\documentclass[twocolumn,jcp]{revtex4-1}
\usepackage[utf8x]{inputenc}
\usepackage{amsbsy,amssymb,amsmath,mathrsfs} 
\usepackage{graphicx} 
\usepackage{color}
\usepackage{listings} 

\newcommand{\dif}{D}

\newcommand{\vect}[1]{\pmb{#1}}
\newcommand{\pf}[0]{\vect{G}}
\newcommand{\pfs}[0]{\vect{g}}
\newcommand{\ppi}[0]{j}
\newcommand{\ppj}[0]{k}

\lstMakeShortInline[columns=fixed]!

\begin{document}

\title{Mesoscopic electrohydrodynamic simulations of binary colloidal suspensions}

\date{\today}

\author{Nicolas Rivas}
\affiliation{Forschungszentrum J\"ulich, Helmholtz Institute
Erlangen-N\"urnberg for Renewable Energy (IEK-11), F\"urther Stra\ss e 248,
90429 N\"urnberg, Germany}
\author{Stefan Frijters}
\affiliation{Department of Applied Physics,
Eindhoven University of Technology, PO Box 513, 5600MB Eindhoven, The
Netherlands}
\author{Ignacio Pagonabarraga}
\affiliation{Departament de F\'isica de la Materia Condensada, Universitat de Barcelona, Barcelona 08028, Spain}
\affiliation{Universitat de Barcelona Institute of Complex Systems (UBICS), Universitat de Barcelona, Barcelona 08028, Spain}
\affiliation{CECAM Centre Europ\'een de Calcul Atomique et Mol\'eculaire, \'Ecole Polytechnique F\'ed\'erale de Lausanne (EPFL), Lausanne CH-1015, Switzerland}
\author{Jens Harting}
\affiliation{Forschungszentrum J\"ulich, Helmholtz Institute
Erlangen-N\"urnberg for Renewable Energy (IEK-11), F\"urther Stra\ss e 248,
90429 N\"urnberg, Germany}
\affiliation{Department of Applied Physics,
Eindhoven University of Technology, PO Box 513, 5600MB Eindhoven, The
Netherlands}

\begin{abstract}
A model is presented for the solution of electrokinetic phenomena of colloidal suspensions in fluid mixtures. We solve the discrete Boltzmann equation with a BGK collision operator using the lattice Boltzmann method to simulate binary fluid flows. Solvent-solvent and solvent-solute interactions are implemented using a pseudopotential model. The Nernst-Planck equation, describing the kinetics of dissolved ion species, is solved using a finite difference discretization based on the link-flux method. The colloids are resolved on the lattice and coupled to the hydrodynamics and electrokinetics through appropriate boundary conditions. We present the first full integration of these three elements. The model is validated by comparing with known analytic solutions of ionic distributions at fluid interfaces, dielectric droplet deformations and the electrophoretic mobility of colloidal suspensions. Its possibilities are explored by considering various physical systems, such as breakup of charged and neutral droplets and colloidal dynamics at either planar or spherical fluid interfaces.
\end{abstract}

\maketitle




\section{Introduction}

Electrokinetic phenomena play a fundamental role in both technological and natural systems, from micro- and nanofluidic devices to molecular biological processes~\cite{li:2004,wong:2004,viovy:2000}. Their physical description has proven challenging mainly due to the wide range of relevant length-scales involved, from interfacial effects, molecular in origin and thus typically on the nanoscale, to the colloidal and system sizes, usually several orders of magnitude larger. An additional significant challenge is to capture the competing effects of two long-range interactions, namely hydrodynamic and electrostatic. These same factors also complicate the realization of experiments as it is, for example, challenging to resolve the charge distribution at solid-fluid or fluid-fluid interfaces, especially at the nanoscale~\cite{siretanu:2014}. In this context, numerical simulations become an attractive alternative to explore those limits where theory and experiments struggle.

Simulations of electrohydrodynamic phenomena vary significantly in methodology, also as a consequence of the wide range of relevant scales. The most common models directly solve the Taylor-Melcher leaky dielectric model, based on the assumption that charge is confined at interfaces in boundaries of negligible length, that is, that the Debye length is small compared to other relevant length-scales~\cite{saville:1997,schnitzer2015taylor}. These are, essentially, macroscopic models and simulations, which ignore the kinetics of the ions and volumetric ionic concentrations. On the opposite side of the spectrum, nanoscale electrokinetic problems, increasing in popularity together with the development of nanofluidics, makes full Molecular Dynamics (MD) simulations a viable alternative for studying electrokinetic phenomena~\cite{liu:2012,chen:2013}. Between these two methods lies a variety of mesoscopic models which permit the simulation of larger time and length-scales than MD, while also resolving the kinetics and spacial distribution of ions. Hydrodynamics can be resolved by mesoscale particle-based methods, such as multi-particle collision dynamics, drastically reducing the number of particles needed compared to full-MD studies~\cite{malevanets:1999,allahyarov:2002,hecht:2005,gompper:2009}. Another common approach involves the solution of Navier-Stokes equations using standard CFD techniques, coupled to Molecular Dynamics for resolving either charged colloids and/or individual ions~\cite{lobaskin:2004}. Full mesoscopic models, which allow the simulation of even longer timescales and a wider range of salt concentrations, treat both the solvents and the ions---through ionic concentrations---at the continuum level. These are usually based on the Nernst-Planck advection-diffusion equation as a model for ion transport, which requires the solution of the Poisson equation to determine the  electric field, comprising a system of equations usually referred to as Poisson-Nernst-Planck theory (PNP)~\cite{capuani:2004}.

In the following we present a numerical solution of an electrohydrodynamic mesoscopic model of colloidal suspensions in binary fluid mixtures with dissolved ions. We focus on nanoscale flows of negligible inertia and comparable electronic, hydrodynamic and colloidal scales. The model considers the ions (solutes) and the fluids (solvents) at the macroscopic level, described by continuous fields of ion concentrations, mass densities and velocity, respectively. The finite-sized colloids are coupled to the fluids and the ions through proper boundary conditions. The simulation methodology recovers the  hydrodynamics by means of the lattice Boltzmann method (LBM)~\cite{benzi:1992}, and the ion kinetics via a finite difference discretization of the Nernst-Planck equation, inspired by the link-flux method~\cite{capuani:2004}. Previous LB-based electrohydrodynamic models for binary mixtures have used a free-energy functional~\cite{evans:1979} to derive the ion kinetics and the forces that the ions exert on the fluids, as well as the interactions between the different fluids~\cite{capuani:2004,rotenberg:2010,pagonabarraga:2010}. Here we present a different approach, deriving the diffusive ion fluxes via the forces exerted on them, and recovering the coupling forces from solvent-solvent and solvent-solute microscopic interactions using pseudopotentials of the form proposed by Shan and Chen~\cite{shan:1993,shan:1994}. The colloidal particles' coupling with the binary mixture is implemented using the Ladd methodology~\cite{ladd:2001}, extended to binary mixtures~\cite{jansen:2011,kruger:2013}. Particles interact between themselves via a Hertz potential when in contact, as also via a lubrication force~\cite{ladd:2001,jansen:2011}. Finally, particle-ion interactions are resolved using a partial-volume discretization that reduces discretization errors, as recently proposed by Kuron \textit{et al.}~\cite{kuron:2016}. Overall this presents the first description of a model capable of handling the electrokinetics of both binary fluid mixtures and moving particles. 

The model is validated by comparing with known theoretical results in several systems encompassing all the method's functionalities. We obtain excellent agreement for dielectric droplet deformation due to Maxwell stresses and distribution of ions at fluid/fluid interfaces. The electrophoretic mobility of colloidal suspensions is also in agreement with previous simulations and experiments. We also show exemplary cases of the possibilities of the model, such as transient dynamics and breakage of droplets in various conditions, colloid wetting properties at fluid/fluid interfaces and the dynamics of colloid-coated droplets. 

The paper is structured as follows: in Section~\ref{sec:model} we first describe the hydrodynamic (\ref{sec:model:hydrodynamics}) and electrokinetic (\ref{sec:model:electrokinetics}) equations, then the interactions between the different species (\ref{sec:model:couplings}), the Poisson's equation (\ref{sec:model:poisson}) and finally the colloidal dynamics (\ref{sec:model:colloids}), each followed by a description of its numerical solution. Section~\ref{sec:results} is dedicated to the study of several systems of increasing complexity, for validation of the model and presentation of the simulation capabilities. Finally conclusions and possible future directions are discussed in Section \ref{sec:conclusions}.






\section{Electrokinetic model} \label{sec:model}

\subsection{Hydrodynamics} \label{sec:model:hydrodynamics}

The lattice Boltzmann method is used to solve two Boltzmann equations with a Bhatnagar-Gross-Krook (BGK) collisional operator~\cite{bhatnagar:1954},
\begin{equation}
	\label{eq:BoltzmannBGK}
	 \left(
		   \frac{\partial}{\partial t} 
		 + \vect{\xi}^\sigma \cdot \nabla_x
		 + \vect{\Phi}_B^\sigma \cdot \nabla_{\vect{\xi}}
	 \right)
	 f^\sigma
	 =
	 \frac{1}{\tau} \left( \tilde
	 f^\sigma - f^\sigma \right),
\end{equation}
where $f^\sigma(\vect{r},\vect{\xi}^\sigma,t)$ is the probability distribution function for component $\sigma \in (a,b)$ with particle velocity $\vect{\xi}^\sigma$; $\tau$ is the relaxation time (the timescale for collisions to drive the local distributions to thermodynamic equilibrium), taken to be equal for both fluids; $\vect{\Phi}^\sigma_B$ the external force and $\tilde f^\sigma(\vect{r},\vect{\xi}^\sigma,t)$ the local equilibrium distribution function. The macroscopic density and the average macroscopic momentum are recovered via $\rho^\sigma = \int f^\sigma \text{d} \vect{\xi}$ and $\rho^\sigma \vect{u}^\sigma = \int \vect{\xi}^\sigma f^\sigma \text{d} \vect{\xi}$, with $\vect{u}^\sigma$ the component macroscopic velocity. When taking the equilibrium distribution as
\begin{align}
	\label{eq:eq_dist}
	\tilde f^\sigma(\vect{x},\vect{v},t)
	= 
	\rho^\sigma \left( \frac{\rho^\sigma}{2\pi p} \right)^{3/2} \exp\left(
	    -\frac{p (\vect{v}^\sigma)^2}{2\rho^\sigma}
    \right),
\end{align}
where $\vect{v}^\sigma = \vect{\xi}^\sigma-\vect{u}$, with $\vect{u}$ the barycentric velocity and $p$ the pressure, it is known that, following the Chapman-Enskog procedure with the Knudsen number as small parameter, the Navier-Stokes equations are recovered, namely
\begin{align}
	\label{eq:continuity}
	& \frac{\partial \rho^\sigma}{\partial t} 
	+ \nabla \cdot \left( \rho^\sigma \vect{u} \right) 
	= 
	0, \\ 
	\label{eq:navier_stokes}
	& \frac{\partial (\rho^\sigma  \vect{u})}{\partial t} 
	+ \nabla \cdot \left( \rho^\sigma \vect{u} \otimes \vect{u} \right)
	= 
	- \nabla \cdot (p \mathbb{I}) 
	+ \nabla \cdot \vect{s}^\sigma
	+ \vect{F}^\sigma.
\end{align}
Here $\otimes$ is the outer product, $p$ the pressure, $\mathbb{I}$ the identity matrix, and $\vect{s}$ the deviatoric stress tensor, $\vect{s}^\sigma = \lambda^\sigma (\nabla \cdot \vect{u}) \mathbb{I} + \eta^\sigma (\nabla \vect{u} + \nabla \vect{u}^T)$, with $\eta$ the dynamic viscosity and $\lambda$ the bulk viscosity. The forcing term $\vect{F}^\sigma$ has two main contributions, coming from interactions between components and electrostatic forces, $\vect{F}^\sigma \equiv \vect{F}_I^\sigma + \vect{F}^\sigma_\text{E}$; both of these terms are derived further down. As the nano-scale is much smaller than the capillary length, gravity is disregarded.

The LBM consists in the discretization of Eq.~\eqref{eq:BoltzmannBGK} in space, time \textit{and} velocities, such that at a given timestep $t$ and at each site of a regular Cartesian lattice $\vect{r}_i$, the distribution function $f^\sigma(\vect{r}, \vect{u}, t)$ becomes $f_d^\sigma(\vect{r}_i,t)$, with $d$ the index of the discrete velocity vectors $\vect{c}_d$~\cite{qian:1992}. Here, we use the usual D3Q19 lattice \cite{qian:1992,benzi:1992} with spacing $\Delta x$, such that particles can travel at each time-step $\Delta t$ to the nearest and next-nearest neighbors. Thus Eq.~\eqref{eq:BoltzmannBGK} can be written as
\begin{multline}
	\label{eq:BoltzmannBGKLattice}
	f^\sigma_d(\vect{r}_i + \vect{c}_d \Delta t, t + \Delta t)
	-
	f^\sigma_d(\vect{r}_i, t) \\
	=
	\frac{\Delta t}{\tau}
	\left(
		\tilde f^\sigma_d(\vect{r}_i, t)
		-
		f^\sigma_d(\vect{r}_i, t)
	\right).
\end{multline}
Notice that we have not included the $\vect{\Phi}^\sigma_B$ term, as external forces will be included as perturbations of the velocity in the equilibrium distribution function $\tilde f^\sigma$, following Shan and Chen~\cite{shan:1993}. After discretization and expansion to second order, Eq.~\eqref{eq:eq_dist} can be written as
\begin{equation}
	\tilde f_d^\sigma
	=
	w_d \rho^\sigma \left[
		1 + \frac{ \vect{c}_d \cdot \vect{u}^\sigma_{e}}{c_s^2} 
	      + \frac{(\vect{c}_d \cdot \vect{u}^\sigma_{e})^2}{2 c_s^4} 
	      - \frac{ \vect{u}^\sigma_e \cdot \vect{u}^\sigma_e}{2 c_s^2}
	    \right],
\end{equation}
The equilibrium velocity of each component is given by $\vect{u}^\sigma_e = \sum_\sigma \rho^\sigma \vect{u}^\sigma / \rho + \tau \vect{\Phi}^\sigma / \rho^\sigma$, with the individual velocities $\vect{u}^\sigma = (1/\rho^\sigma)\sum_d f_d^\sigma \vect{c}_d$. The lattice weights for a D3Q19 lattice are
\begin{equation}
w_d =
\begin{cases}
	1/3 & \text{if }  |\vect{c}_d| = 0, \\
	1/18 & \text{if } |\vect{c}_d| = \Delta x / \Delta t, \\
	1/36 & \text{if } |\vect{c}_d| = \sqrt{2} (\Delta x / \Delta t).
\end{cases}
\end{equation}



\subsection{Electrokinetics} \label{sec:model:electrokinetics}

The cations and anions dispersed throughout the solvents are considered at the continuum level. The evolution of the concentration of each ion species is given by the advection-diffusion Nernst-Planck equation,
\begin{equation}
	\label{eq:nernst-planck}
	  \frac{\partial n^\pm}{\partial t} 
	+ \vect{u} \cdot \nabla n^\pm
	=
	\nabla \cdot \vect{j}^\pm,
\end{equation}
where $n^\pm(\vect{r})$ is the number density of cations ($^+$) and anions ($^-$). The diffusive ion flux is
\begin{equation}
	\label{eq:ionicflux}
	\vect{j}^\pm
	=
	  \dif \nabla n^{\pm} 
	+ \dif \beta \vect{F}^\pm
\end{equation}
with the diffusivity $D$ (which we have assumed to be homogeneous and the same for both species), and $\beta \equiv 1/k_B T$, with the Boltzmann constant $k_B$ and the temperature $T$. Notice that this temperature only serves as an energy scale, as we have considered isothermal systems with no fluctuations. In all our studies we take $\beta = 1$, although for clarity we keep the symbol $\beta$ where present.

The Nernst-Planck equation ignores ion-ion interactions. These are known to be relevant at the nanoscale for moderate ion concentrations, where steric and electric interactions can alter the flow of ions~\cite{greberg1998charge,hunter2003significance,qiao2004charge,van2006charge}. Therefore our model is strictly valid for low ionic concentrations $n^\pm(\vect{r}) \ll 1/\Delta x^3$.

The force density applied to the ions, $\vect{F}^\pm(\vect{r})$, is coupled to the flux through the Smoluchowski relation with mobility $D\beta$. This implies that the inertial timescale of the ions is much faster than any other resolved timescale. Therefore ions instantly reach their drift velocity as $\vect{F}^\pm(\vect{r})$ varies, their dynamics being dominated by the solvents' viscosity. The force has two contributions, electrostatic and a term coming from the microscopic interactions between solvents and solutes,
\begin{equation}
	\label{eq:electric_ion_force}
	\vect{F}^\pm = q^\pm \nabla \phi - \vect{F}_I^\pm,
\end{equation}
with $q^\pm(\vect{r}) = z^\pm e n^{\pm}(\vect{r})$, $z^\pm$ the valence of cations and ions and $-e$ the electron charge. 
In simulations we take $e$ as the unit of charge, $e = 1$, although for clarity we keep the symbol in the expressions. Furthermore, we only consider monovalent salts, $z^\pm = \pm 1$. 

The total electric potential $\phi(\vect{r})$ is the sum of an internal and external contribution $\phi(\vect{r}) = \phi_I(\vect{r}) + \phi_E(\vect{r})$, with the external one usually being a parameter used to control the flow. Analogously, we define equivalent electric fields as $\vect{E}(\vect{r}) = - \nabla \phi(\vect{r}) = \vect{E}_I(\vect{r}) + \vect{E}_E(\vect{r})$.
The internal contribution is itself a function of the charge distribution via Poisson's equation,
\begin{align}
	\label{eq:electric_potential}
  & \nabla \cdot (\varepsilon \nabla \phi_I)
  =
  - q,
\end{align}
where we have defined the total charge density $q(\vect{r}) \equiv e(z^+ n^+(\vect{r}) + z^- n^-(\vect{r})) + q_p(\vect{r})$. Here $q_p(\vect{r})$ is the total charge from suspended particles, as will be detailed further below.

The permittivity $\varepsilon(\vect{r}) = \varepsilon_0 \varepsilon_r(\vect{r})$, with $\varepsilon_0$ the vacuum permittivity and $\varepsilon_r(\vect{r})$ the relative permittivity of the solution, is not homogeneous, as each solvent and the colloidal particles can have different permittivities. At each point, $\varepsilon(\vect{r})$ is either a function of the fluid concentration, or the permittivity of the particles, such that
\begin{equation}
	\varepsilon(\vect{r}) = 
	\begin{cases}
		\varepsilon^c(\vect{r}) & \vect{r} \text{ inside the particle},\\
		\varepsilon^f(\vect{r}) & \vect{r} \text{ inside the fluid}.
	\end{cases}
\end{equation}
As the dependency of the fluid's permittivity on the concentration of each fluid component is a priori unknown, we follow Rotenberg \textit{et al.}~\cite{rotenberg:2010} in taking the simplest model, a linear interpolation between the value of each fluid, $\varepsilon_a$ and $\varepsilon_b$, as a function of the local concentration,
\begin{equation}
	\label{eq:dielectric_constant}
	\varepsilon^f(\vect{r})
  = \tfrac{1}{2} \left(
  		\varepsilon^a (1-c(\vect{r}))
	  + \varepsilon^b(1+c(\vect{r})) \right),
\end{equation}
with $c(\vect{r}) \equiv (\rho^b(\vect{r}) - \rho^a(\vect{r}))/(\rho^a(\vect{r}) + \rho^b(\vect{r}))$.

The Nernst-Planck equation \eqref{eq:nernst-planck} is numerically solved using a finite difference scheme with different discretization methods for the advective and diffusive terms, inspired by the link-flux method~\cite{capuani:2004,rotenberg:2010}. Integrating Eq.~\eqref{eq:nernst-planck} over a lattice site cell's volume, which we assume to be always a cube of side length $\Delta x$, results in
\begin{multline}
	\label{eq:nernst-planck-disc}
	\frac{n^\pm(\vect{r}_i, t + \Delta t)- n^\pm(\vect{r}_i, t)}{\Delta t}
	= 
	  \frac{1}{\Delta x} \times \\
	  \sum_{d \in \{Q6\}} \left[ - \vect{u}(\vect{r}_i) n^\pm (\vect{r}_{i+d/2})
	+ \vect{j}^\pm (\vect{r}_{i+d/2})
	\right]\cdot \vect{c}_{d}.
\end{multline}
where we have defined $\vect{r}_{i+d/2} \equiv \vect{r}_i + \vect{c}_{d}\Delta t/2$. Here we take $d$ to run only over the D3Q7 lattice directions, as it simplifies the discretization and improves performance. Note that in the D3Q7 lattice $\vect{c}_{d}$ is equivalent to the normal of the surface of the lattice cell in the $d$ direction, which is always a square of area $\Delta x^2$. Previous research using the link-flux method has always considered a D3Q19 lattice, both to have consistent discretizations for solutes and solvents and to reduce the spurious diffusion produced by advection. Using the methodology detailed in Ref.~\cite{capuani:2004}, we have measured that in our case the effective diffusivity stays well below a $2\%$ difference with the nominal diffusivity $D$, for the comparably low value used here $D = 0.01$, and the largest possible fluid velocities. Previous researchers have performed careful comparisons of the accuracy of different differential operators, revealing that higher lattice connectivities indeed increase the accuracy and convergence (as a function of the resolution), although total errors are expected to be negligible in our cases of variations on the order of the diffusive interface~\cite{ramadugu2013lattice}.

For the advection terms we use a first-order upwind discretization, such that
\begin{multline}
	\sum_{d\in\{Q6\}} n^\pm (\vect{r}_{i+d/2}) \vect{u}(\vect{r}_i) \cdot \vect{c}_{d}
	= \\
	\sum_{d\in\{Q6\}} H[\vect{u}(\vect{r}_i) \cdot \vect{c}_{d}] n^\pm (\vect{r}_i) \vect{u}(\vect{r}_i) \cdot \vect{c}_{d} ,
\end{multline}
with $H[x]$ the Heaviside function. For the diffusive flux we follow a different procedure than in Refs.~\cite{capuani:2004,rotenberg:2010}, as it has been recently shown that their proposed exponentiation of the fluxes leads to a quadratic error for flows far from equilibrium~\cite{rempfer:2016}. We thus use a straightforward finite-difference discretization for the gradients, and a linear interpolation to obtain the value of the fields at the volumes' surfaces. This has been shown to be more computationally efficient while also reducing spurious fluxes~\cite{rempfer:2016}. Denoting $\vect{r}_{i+d} \equiv \vect{r}_i + \vect{c}_{d} \Delta t$, Eq.~\eqref{eq:ionicflux} results in
\begin{multline}
	\label{eq:ionic-fluxes-disc}
	\vect{j}^\pm(\vect{r}_{i+d/2}) 
	= 
	\dif
	\left[
		\left(
		    \frac{n^{\pm}(\vect{r}_{i+d})-n^{\pm}(\vect{r}_i)}{\Delta x}
		\right) \vect{c}_{d}
		\right.\\
		\left.
	  - \beta 
		\left(
		    \frac{q^{\pm}(\vect{r}_i)+q^{\pm}(\vect{r}_{i+d})}{2}
		\right)
		\left(
		    \frac{\vect{E}(\vect{r}_i)+\vect{E}(\vect{r}_{i+d})}{2}
		\right)
		\right.\\
		\left.
		-
		\vect{F}_I^\pm(\vect{r}_{i+d/2})
	\right].
\end{multline}
The form and discretization of the last term, the interaction force, is given in what follows.

\subsection{Couplings} \label{sec:model:couplings}

Microscopic interactions between the different solvents and solutes are modeled using the pseudopotential method of Shan and Chen~\cite{shan:1993}, which is among the most popular multiphase/multicomponent LB methods~\cite{liu:2016}. It is usually preferred to other alternatives due to its simplicity of implementation. The method allows us to determine $\vect{F}_I^\pm(\vect{r})$ and $\vect{F}_I^\sigma(\vect{r})$ in a consistent manner. Microscopic molecular interactions between species (either solvents or solutes) are captured as local force densities of the form
\begin{equation}
	\label{eq:shanChenForce}
	  \vect{\Phi}_I^\alpha(\vect{r}_i\,) 
	= 
	- \psi^\alpha(\vect{r}_i\,) 
	  \sum_{\beta \neq \alpha} 
	  \sum_{d} 
	  G_{\alpha \beta} w_d \psi^{\beta} (\vect{r}_{i+d}) \vect{c}_d \Delta t,
\end{equation}
with the indexes $\alpha,\beta \in \{a,b,+,-\}$, $\alpha \ne \beta$. As we have neglected ion/ion interactions, we take the coupling constants $G_{+-} = G_{-+} = 0$

Let us first focus on fluid/fluid interactions, $\alpha, \beta \in \{a,b\}$. The coupling constant $G \equiv G_{ab} = G_{ba}$ determines the strength of attraction ($G < 0$) or repulsion ($G > 0$) between the solvents. The pseudopotential $\psi^\sigma(\vect{r}_i,t)$ is taken to have the common form
\begin{equation}
	\psi^\sigma(\vect{r}_i\,) = \rho_0 \left[ 1 - \exp(-\rho^\sigma(\vect{r}_i\,)/\rho_0) \right].
\end{equation}
The reference density is set to $\rho_0 = 1$, although we leave the symbol for clarity. In the continuum limit, and to fourth order in derivatives, the Shan-Chen force \eqref{eq:shanChenForce} can be shown to be~\cite{sbragaglia2009continuum}
\begin{equation}
    \vect{F}_{If}^\sigma = - G \psi^\sigma
    \left(
        c_s^2 \Delta t^2 \nabla \psi^{\tilde \sigma} + 
        \frac{c_s^4 \Delta t^4}{2} \nabla \left( \Delta \psi^{\tilde \sigma} \right) 
    \right).
\end{equation}
This results in a modified equation of state~\cite{shan:1993}
\begin{equation}
	\label{eq:eqofstate}
	p 
	= 
	c_s^2 \rho 
	+ 
	c_s^2 \Delta t^2 G \psi^a \psi^b,
\end{equation}
with the speed of sound $c_s = \Delta x/\sqrt{3} \Delta t$. It can thus be seen that, depending on the value of $G$, Eq.~\eqref{eq:eqofstate} leads to phase separation. In this study we always take $G = 4.0\Delta t^{-2} \rho_0^{-1}$.

Next, we consider the case of fluid/ion microscopic interactions, referred to as solvation. These are the cases $\alpha\in\{a,b\}$, $\beta\in\{+,-\}$ in Eq.~\eqref{eq:shanChenForce}. For the ionic pseudopotentials we take the same form as for the fluids, although as $n^\pm(\vect{r}) \ll 1/\Delta x^3$ they can be simplified,
\begin{equation}
    \psi^\pm(\vect{r}_i) = n^\pm_0(1 - \exp(-n^\pm(\vect{r}_i)/n^\pm_0)) \approx  n^\pm(\vect{r}_i).
\end{equation}
with the reference density $n_0^\pm = 1/\Delta x^3$. It follows that, in the continuum limit, the forces applied to the fluids from ion interactions take the form
\begin{equation}
    \label{eq:solvation:force1}
    \vect{F}^\sigma_{Ii} = - \sum_\pm G_{a\pm} \psi^a c_s^2\Delta t^2 \nabla n^\pm.
\end{equation}
We have here assumed, for simplicity, that $G_{b\pm} = 0$. This is without loss of generality, as the coupling coefficients can also be negative, and thus hold the possibility of modelling both attraction and repulsive forces from solvent $a$, the latter equivalent in our scheme to an attractive force to solvent $b$. 

Analogously, the forces applied to the ions from their microscopic interaction with the fluids take the form
\begin{equation}
    \label{eq:solvation:force2}
    \vect{F}^\pm_{I} = -G_{a\pm} c_s^2\Delta t^2 n^\pm \nabla \psi^a.
\end{equation}

In order to specify the value of $G_{a\pm}$, we interpret our model of ion/solute microscopic interactions as an approximate model of solvation. Solvation interactions are captured at the macroscopic level by chemical potentials, $\mu^\pm_\text{s}(\vect{r})$, corresponding to the cost in free energy of adding an ion to the solvent or mixture. For example, in the most common case of hydration (solvation in water), $\mu^\pm_\text{s}(\vect{r})$ is fundamentally a function of the hydrogen bonds formed between the ions and the solvent molecules~\cite{onuki:2006,onuki2011phase}. Selective solvation in mixtures, when there is a difference in the solvation energies between the two solvents and solutes, can have crucial effects on the global dynamics, as usually the associated energies far exceed $k_B T$~\cite{onuki2011phase}. Even though the solvation energies are known to depend on several parameters, here we follow a similar approach as Onuki \textit{et al}.~\cite{onuki2011phase} and take a simple form of a linear dependency with the density of solvent $a$,
\begin{equation}
	\label{eq:solvation_potential}
	\mu^\pm_\text{s}(\rho^a)
	= 
    - \frac{\Delta \mu^\pm}{\Delta \rho^a} \rho^a(\vect{r}).
\end{equation}
The parameter $\Delta \mu^\pm = \mu^{\pm}_\text{s}(\bar \rho^a_a) - \mu^{\pm}_\text{s}(\bar \rho^a_b)$, commonly referred to as Gibbs transfer energy~\cite{onuki:2006,rotenberg:2010}, is the difference of the respective solvation chemical potentials at the bulk of each component in a phase-separated system. For example, in the case of a planar interface at $x=0$ between two immiscible fluids, $\bar \rho^{a}_a = \rho^a(-\infty)$, and $\bar \rho^{a}_b = \rho^a(\infty)$. Analogously $\Delta \rho^a \equiv \bar \rho^a_a - \bar \rho^a_b$.

Variations of chemical potentials translate to forces, given by the Gibbs-Duhem relation (at constant temperature) $\vect{f} = -\nabla p = \rho \nabla \mu_\rho$. Therefore, variations in the solvation chemical potential generate a force of the form $-(\Delta \mu^\pm / \Delta \rho^a) n^\pm(\vect{r}) \nabla \rho^a(\vect{r})$. Comparing with Eq.~\eqref{eq:solvation:force2}, and noticing that for small densities $\psi^a \approx \rho^a$, it is evident that in order to interpret the pseudopotential interactions as solvation effects, the Shan-Chen force coupling parameter has to be taken as
\begin{equation}
    G_{a\pm} = \frac{\Delta \mu^\pm}{\Delta \rho^a c_s^2 \Delta t^2}.
\end{equation}

Having determined the forces stemming from microscopic molecular interactions between solvents and solutes, two other forces rest to be considered. The first comes from the relative movement of the ions with the respect to the solvent, and is given by a simple friction coupling between the two, that is 
\begin{equation}
    \label{eq:frictionforce}
    \vect{F}_{Ij}^\sigma = - \frac{k_B T}{D} \sum_\pm \vect{j}^\pm.
\end{equation}

The second remaining force has its origin on the dielectric nature of the solvents. Polarization effects give rise to the Kelvin force density, which assuming an isotropic permittivity takes the form $\vect{F}_{E}(\vect{r}) = - \tfrac{1}{2} E(\vect{r})^2 \nabla \varepsilon + \tfrac{1}{2} \nabla (E(\vect{r})^2 \phi \partial \varepsilon / \partial \phi)$~\cite{landau1960electrodynamics}. Using Eq.~\eqref{eq:dielectric_constant} gives
\begin{equation}
    \label{eq:kelvin}
    \vect{F}_{E}(\vect{r}) 
    = - \tfrac{1}{2} (\varepsilon(\vect{r}) - \bar \varepsilon) \nabla E(\vect{r})^2,
\end{equation}
with the average permittivity $\bar \varepsilon = \tfrac{1}{2}(\varepsilon^a + \varepsilon^b)$. The Kelvin force density is strictly valid in the electroquasistatic limit and for incompressible media~\cite{zahn:2006}. The latter condition is fulfilled considering flows of low Mach number, $\text{Ma} = u_0 / c_s \ll 1$, with $u_0$ the typical flow speed. The electroquasistatic limit is satisfied for small ion fluxes, that is, $D \ll \Delta x^2 / \Delta t$. In what follows we always take $D = 10^{-2} \Delta x^2/\Delta t$.

Notice that in Eq.~\eqref{eq:kelvin} we have disregarded the electrostatic term of the Kelvin force density, as it has already been included in Eq.~\eqref{eq:frictionforce}. This is just a consequence of the macroscopic Lorentz force being simply the sum of the individual ion contributions.

In summary, the external force term in the Navier-Stokes equation \eqref{eq:navier_stokes} is given by
\begin{equation}
    \label{eq:totalhydroforce}
    \vect{F}^\sigma
    = 
      \vect{F}^\sigma_{If}
    + \vect{F}^\sigma_{Ii}
    + \vect{F}^\sigma_{Ij}
    + \vect{F}^\sigma_E.
\end{equation}
After some manipulation, and defining the total ionic concentration $n(\vect{r}) \equiv n^+(\vect{r}) + n^-(\vect{r})$, Eq.~\eqref{eq:totalhydroforce} can be written as
\begin{multline}
	\label{eq:electric_fluid_force}
	\vect{F}^\sigma 
	= 
	  \vect{F}^\sigma_{If} \\
	- \frac{\nabla n}{\beta}
	- \sum_\pm \nabla
	  \left(
	    n^\pm \mu_\text{s}^\pm 
	  \right)
	+ q \vect{E}
	- \frac{(\varepsilon - \bar \varepsilon)}{2} \nabla E^2.
\end{multline}

Finally we present the discretization of the force terms. The forces on the solvent due to microscopic interactions are given by Eq.~\eqref{eq:shanChenForce}, with the already specified coupling constants and pseudopotentials. For Eq.~\eqref{eq:frictionforce} we use the discretization of the fluxes, Eq.~\eqref{eq:ionic-fluxes-disc}, with
\begin{multline}
    \vect{F}^\pm_I(\vect{r}_{i+d/2}) = \\
    - \frac{\Delta \mu^\pm}{\Delta \rho^a}
	\left(
	    \frac{n^\pm(\vect{r}_i)+n^\pm(\vect{r}_{i+d})}{2}
	\right)
	\left(
		\frac{\rho^a(\vect{r}_{i+d})-\rho^a(\vect{r}_i)}{\Delta x}
	\right)
	\vect{c}_{d}.
\end{multline}
Therefore, the total force included in the lattice Boltzmann method is given by
\begin{multline}
    \vect{\Phi}^\sigma (\vect{r}_i)
    = 
    \vect{\Phi}^\sigma_I(\vect{r}_i)
    + \\
    \sum_{d\in\{Q6\}} \vect{F}_{Ij}^\sigma(\vect{r}_{i+d/2})  
    +
    \left(\frac{\varepsilon(\vect{r}) - \bar \varepsilon}{4}\right) E^2(\vect{r}_n) \vect{c}_{d}.
\end{multline}

\subsection{Poisson equation} \label{sec:model:poisson}

A significant challenge in the numerical implementation of electrokinetic models is the efficient and accurate solution of Poisson's equation~\eqref{eq:electric_potential}. $\phi_\text{I}(\vect{r})$ must be determined at each time-step from the charge distribution $q(\vect{r})$ and the permittivity $\varepsilon(\vect{r})$. Here we use a standard Fourier spectral method, which allows us to find a solution for the potential in $\mathcal{O}(G\log(G))$, where G is the number of lattice sites. A significant advantage of this method is the possibility of efficient parallelization. On the other hand, boundary conditions on the electric potential can be hard to implement, although here we consider only periodic systems.

One additional difficulty in our case comes from the space dependent $\varepsilon(\vect{r})$. In that case, the general Poisson equation~\eqref{eq:electric_potential} has to be solved, which can be written as
\begin{equation}
	\nabla^2 \phi_I(\vect{r}, t) = - q(\vect{r},t)/\varepsilon(\vect{r},t) - \nabla \varepsilon(\vect{r},t) \cdot \nabla
	\phi_I(\vect{r}, t).
\end{equation}
As $\varepsilon(\vect{r},t)$ is known, only the $\nabla\phi$ term on the right presents a problem. In order to decouple this term and be able to use the standard discretization of the Poisson equation, we assume a slowly varying potential, such that $\phi_I(t) \sim \phi_I(t-\Delta t)$, and thus
\begin{equation}
	\nabla^2 \phi_I(\vect{r}, t) = - q(\vect{r}, t)/\varepsilon(\vect{r},t) - \nabla \varepsilon(\vect{r},t) \cdot \nabla 	
	\phi_I(\vect{r}, t-\Delta t).
\end{equation}

The condition of a slowly varying $\phi_I(\vect{r},t)$ is essentially a demand for a slowly varying $q(\vect{r}, t)$, a condition already set by the total electric force by assuming low Mach numbers and diffusivities, which imply small advection and diffusion of ions, respectively.


\subsection{Colloidal dynamics} \label{sec:model:colloids}

Having specified the model and numerical implementation for the hydrodynamics and ion kinetics, only the dynamics of the colloidal particles rests to be determined. For simplicity, we consider only spherical and rigid particles of radius $r_p$. Particles follow Newton's equation of motion
\begin{align}
	\vect{\pf}^\ppi
	= 
	m \frac{d^2\vect{x}^\ppi}{dt^2},
\end{align}
with $\ppi = 1,..., N$ the particle index, $N$ the total number of particles, $m$ the mass, $\vect{x}^\ppi$ the position and $\vect{\pf}^\ppi$ the total force exerted on particle $\ppi$. For simplicity we take all particles to have the same mass $m = \rho_p (4/3) \pi r_p^3$, with $\rho_p = 5\rho_0$, an arbitrary choice which has been seen to have no influence on the results. The total force includes both particle-particle and particle-fluid interactions, as well as an electrostatic term,
\begin{align}
	\vect{\pf}^\ppi
  = 
    \vect{\pf}_\text{fl}^\ppi
  + \vect{\pf}_\text{lub}^\ppi
  - Q^\ppi \nabla \phi 
  - \sum_j \nabla V(\vect{x}^\ppi,\vect{x}^\ppj),
\end{align}
with $V(\vect{x}^\ppi,\vect{x}^\ppj)$ the interaction potential between particles $\ppi$ and $\ppj$, $\vect{\pf}_\text{fl}^\ppi$ the force exerted on the particle by the fluids and $\vect{\pf}_\text{lub}^\ppi$ the lubrication correction. Also present is the force exerted by the electric field on the particle, which is assumed to have a charge $Q^\ppi$ homogeneously distributed throughout its volume.

The interaction potential is given by the Hertzian model
\begin{equation}
    \label{eq:colloid_interaction}
	\nabla V(\vect{x}^\ppi, \vect{x}^\ppj) = k_H \delta^{3/2} \hat{\vect{n}}^{\ppi\ppj},
\end{equation}
with the direction between particles' centers $\hat{\vect{n}}^{\ppi\ppj} = (\vect{x}^\ppi - \vect{x}^\ppj)/||\vect{x}^\ppi - \vect{x}^\ppj||$, and $\delta$ the overlap between the two particles, $\delta = \max\{2r_p - ||\vect{r}^\ppi - \vect{r}^\ppj||,0\}$. We take the stiffness to be $k_H = 1.0 (m_0 \Delta t^{-2} \Delta x^{-1/2})$, although we see no influence of its value in any of our considered system due to the small overlaps and colloid volume fractions involved.

The lubrication force $\vect{\pf}_\text{lub}^\ppi$ models an observed repulsive interaction between two particles approaching each other in a fluid medium~\cite{ladd:2001}. It has its origin in strong pressure gradients generated by the flow of fluid out of the gap between the particles as they approach each other. Here we follow the procedure of Ladd \textit{et al.}~\cite{ladd:2001}, including lubrication effects directly as a force of the form
\begin{multline}
    \label{eq:lubrication}
	\vect{\pf}_\text{lub}^\ppi 
	= 
	- \sum_\ppj \frac{3\pi}{2} \eta^\sigma r_p^2\hat{\vect{n}}^{\ppi\ppj}
	\left[ \hat{\vect{n}}^{\ppi\ppj} \cdot (\vect{v}^\ppi - \vect{v}^\ppj) \right] \\
	\times \left[(||\vect{x}^\ppi - \vect{x}^\ppj|| - 2 r_p)^{-1} - 1\right],
\end{multline}
with the particle velocities $\vect{v}^\ppi = d\vect{x}^\ppi/dt$, and the sum over $\ppj$ runs over all particles such that $\ppj\ne\ppi$ and $||\vect{x}^\ppi - \vect{x}^\ppj|| - 2 r_p < \tfrac{2}{3}\Delta x$.

These two forces provide a simplified description of what are in reality highly complex colloid/colloid interactions. A more involved potential than Eq.~\eqref{eq:colloid_interaction} would be needed to model the interactions at small distances, where van der Waals and other intermolecular interactions are relevant. The derivation of the correct interaction potential at these scales is not trivial. For example, the commonly used DLVO potential is not directly applicable, as the electrostatic interaction between Debye layers is expected to be already resolved at large distances, although not when the separation between the colloids becomes small, $||r^j - r^k|| < \Delta x$. Moreover, a more general description is needed for colloids at fluid interfaces, a topic of active research~\cite{oettel2005attractions,dominguez2008multipole,majee2014electrostatic,girotto2015interaction,majee2016poisson}. Finding a correction for electrostatic interactions, such as Eq.~\eqref{eq:lubrication}, is suggested as a future direction of research.

\begin{figure} 
	\begin{center}
		\includegraphics[scale=0.59]{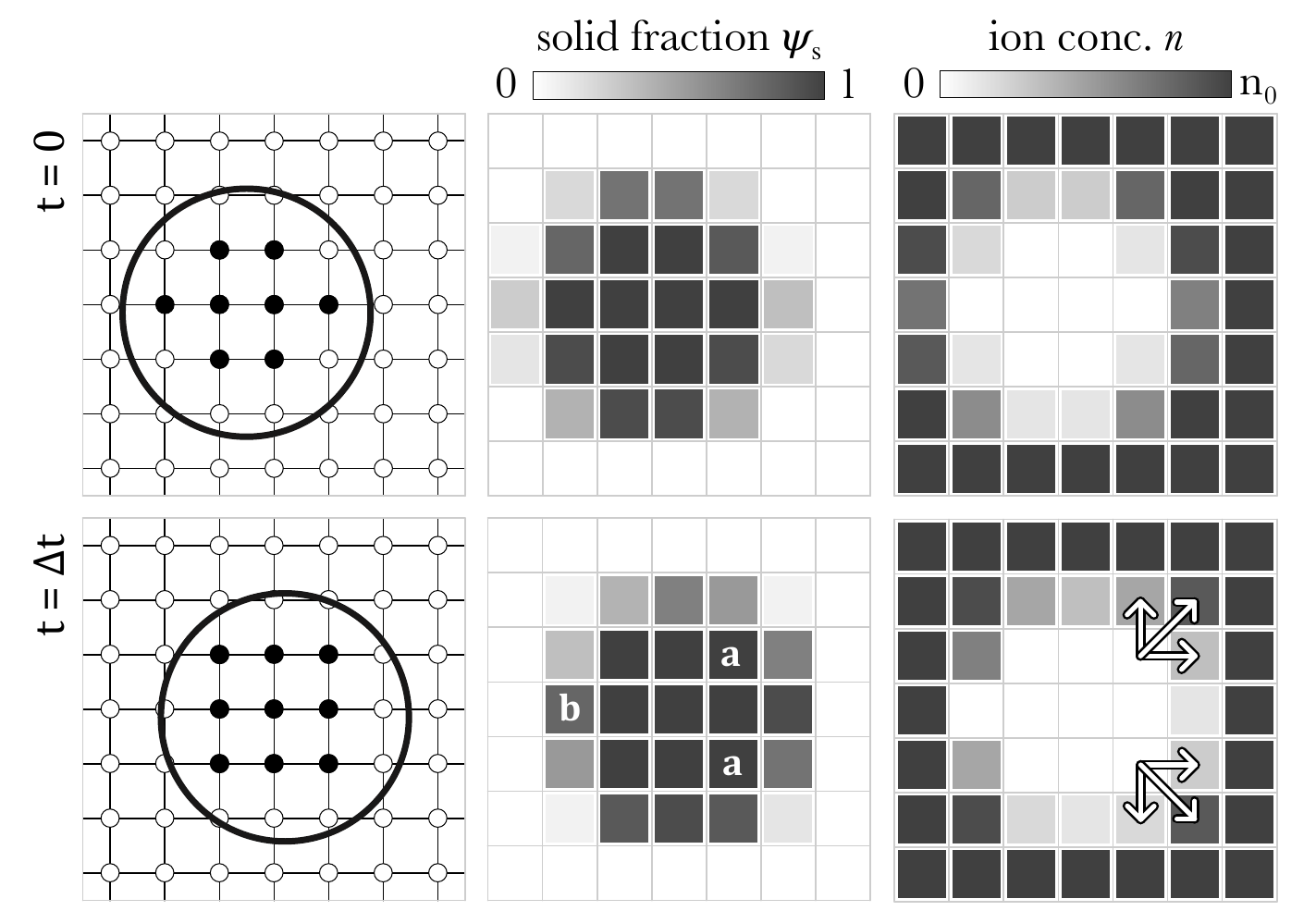}
	\end{center}
	\caption{Scheme of the projection of a particle, presented in 2D for simplicity. The first column shows the perimeter of the particle and the lattice sites, with white for fluid and black for solid sites. The second and third columns show the solid fraction $\psi(\vect{r}_i)$ and the ion concentration $n(\vect{r}_i)$, respectively. The two rows show successive snapshots in time. Created and deleted solid sites are marked with $a$ and $b$, respectively, and the corresponding displacement of ions is indicated with arrows. }
	\label{fig:projection}
\end{figure}

Finally, only the force resulting from the interaction between fluids and colloids rests to be specified, $\vect{\pf}_\text{fl}^\ppi$. This is resolved by projecting the particles onto the lattice, marking as solid the sites that are covered by the particle. The interaction is then a consequence of the boundary conditions applied to fluid advecting towards particle solid nodes. Furthermore, as the particle moves, two other processes need to be specified: the conversion of fluid sites to solid, and the conversion of solid sites to fluid. Fig.~\ref{fig:projection} shows an illustration of these processes.

Fluid advecting onto particle solid sites is bounced-back half way between the solid and the fluid sites, by setting
\begin{equation}
	f_d^\sigma(\vect{r}_i,t) = f_{d^*}^\sigma(\vect{r}_i,t+\Delta t) - \frac{1}{6} \rho^\sigma(\vect{r}_i,t) \vect{v}^\ppi \cdot \vect{c}_{d^*} \frac{\Delta t^2}{\Delta x^2}.
\end{equation}
where $\vect{c}_d = - \vect{c}_{d^*}$. The second term corresponds to a correction that takes into account the particle movement~\cite{aidun:1998,ladd:2001,jansen:2011}. The force density transferred to the particle from every bounce-back collision is then
\begin{equation}
	\vect{\pfs}^\ppi_\text{bb}(\vect{r}_i)
	= 
	\frac{1}{\Delta t} 
	\left( 
		2 \rho^\sigma(\vect{r}_i) 
		- \frac{1}{6} \rho^\sigma(\vect{r}_i) \vect{v}^\ppi \cdot \vect{c}_{d^*} \frac{\Delta t^2}{\Delta x^2}
	\right) \vect{c}_{d^*}.
\end{equation}

When creating solid sites, as the particle moves and occupies a new lattice site, the momentum of the fluid being covered must be transferred to the particle, in order to ensure global conservation of momentum. Therefore an additional term has to be added to the particle,
\begin{equation}
	\vect{\pfs}^\ppi_\text{c} = - \frac{\rho^\sigma (\vect{r}_i) \vect{u}(\vect{r}_i)}{\Delta t}.
\end{equation}
Moreover, in the last case of a particle uncovering a solid site, new fluid is created from an equilibrium distribution with the velocity of the particle at that site, such that
\begin{equation}
	f_d^\sigma(\vect{r}_i,t)
	=
	\bar\rho^\sigma 
	\tilde f_c^\sigma (\vect{v}^\ppi(\vect{r}_i,t),\rho(\vect{r}_i,t)).
\end{equation}
The newly added density $\bar\rho^\sigma$ corresponds to the average of the neighboring sites,
\begin{equation}
	\bar \rho^\sigma = \sum_{d} \rho^\sigma(\vect{r}_{i+d})
\end{equation}
where the sum runs over the sites' neighbors. Correspondingly, the final contribution to the total force on the particle is
\begin{equation}
	\vect{\pfs}^\ppi_\text{d} = \frac{\bar \rho^\sigma \vect{u}(\vect{r}_i,t)}{\Delta t}.
\end{equation}
The total force from fluid interactions is then obtained by summing all individual contributions at every timestep, such that $\pf_\text{fl}^\ppi = (\sum_\ppi \vect{\pfs}^\ppi_\text{bb}(\vect{r}_i) + \pfs_\text{c}^j(\vect{r}_i) + \pfs_\text{d}^\ppi(\vect{r}_i))\Delta x^3$, with the sum running through all relevant lattice sites of each process.

A final correction due to the presence of colloids in binary fluids involves the computation of Shan-Chen forces. As can be seen in Eq.~\eqref{eq:shanChenForce}, the density of the neighbors of a given site need to be accessed to determine the pseudopotentials, but right next to the particle these might be inside the particle. When solid sites are just considered to have no fluid the result is an artificial increase in the density next to the particle~\cite{jansen:2011}. To solve this problem we use the methodology proposed by Jansen~\textit{et al}~\cite{jansen:2011}, and set the interface sites inside the particle to have a virtual fluid concentration equal to the average of the fluid neighboring sites, $\bar \rho^\sigma$. No advection or collision steps are performed on the virtual fluid sites. Nonetheless, these are considered for Shan-Chen forces, balancing the forces at the fluid side and thus preventing the formation of a high density layer~\cite{jansen:2011}.

For the boundary conditions of the ion fluxes with the colloids we follow an equivalent procedure as proposed by Kuron~\textit{et al.}~\cite{kuron:2016}. As in general an electrical double layer is formed around the colloids, where ion concentrations are considerably larger than in the bulk, the displacement of charges due to the creation and removal of solid sites leads to sudden electric field variations, which result in large fluctuations of the particles' velocities. In order to reduce these discretization effects, the flux of ions is modified to take into account the volume fraction covered by the particle at each site. The overall procedure was inspired by Noble-Torczynski hydrodynamic boundary conditions, which generalize the previously described bounce-back procedure for sites that are partially filled by a solid~\cite{noble:1998}. The solid fraction $\psi_s(\vect{r}_i)$ is determined at every lattice site, with $\psi_s = 0$ for lattice site volumes that do not intersect with any particle, and $\psi_s = 1$ for volumes completely inside a particle, that is, solid sites (see Fig~\ref{fig:projection}). For the cases of partial overlap, the solid fraction is exactly given by the intersection of a sphere with a cube of side length $\Delta x$ at the position of the site $\vect{r}_i$. As there is no simple analytical solution for this geometrical problem, $\psi$ is approximated by subdividing each lattice site $h$-times, and counting the number $\nu$ of cubic central points that are inside the spherical colloid. It then follows that $\psi_s = \nu / (h+1)^3$.

The ionic fluxes (Eq.~\eqref{eq:ionic-fluxes-disc}) are modified to take into account the solid fraction field $\psi_s$, such that at equilibrium the amount of ions at a site is proportional to the fluid fraction, $\psi_f = 1 - \psi_s$. The fluxes then take the form
\begin{multline}
	\label{eq:ionic-fluxes-disc-2}
	\vect{j}^\pm(\vect{r}_{i+d/2})
	= 
	-
	\dif
	\left[
		\frac{1}{\Delta x}
		\left( 
		\frac{n^{\pm}(\vect{r}_{i+d})}{\psi_f(\vect{r}_{i+d})} 
		-
		\frac{n^{\pm}(\vect{r}_i)}{\psi_f(\vect{r}_i)} 
		\right) \vect{c}_{d}
		\right.\\
		- 
		\left.
		\frac{\beta}{4}
		\left(
			\frac{q^{\pm}(\vect{r}_i)}{\psi_f(\vect{r}_{i})} 
			+
			\frac{q^{\pm}(\vect{r}_{i+d})}{\psi_f(\vect{r}_{i+d})}
		\right)
		\left(
			\vect{E}(\vect{r}_i)
			+
			\vect{E}(\vect{r}'_i)
		\right)
		\right.\\
		\left.
		+
		\frac{\beta \Delta \mu^\pm}{2 \Delta \rho^a \Delta x}
		\left(
			\frac{n^\pm(\vect{r}_{i+d})}{\psi_f(\vect{r}_{i+d})}+\frac{n^\pm(\vect{r}_i)}{\psi_f(\vect{r}_i)}
		\right)
		\left(
		    \rho^a(\vect{r}_{i+d})-\rho^a(\vect{r}_i)
		\right)\vect{c}_{d}
	\right] \\ \times 
	\psi_f(\vect{r}_{i+d})  \psi_f(\vect{r}_i).
\end{multline} 

The creation and deletion of solid sites has to also take into account the ions present in the fluid. At the creation of a solid site, ions are isotropically displaced to the fluid neighboring sites. Although this might seem artificial, it is always a relatively small amount of ions, as $\psi_s \sim 1$ (see Fig.~\ref{fig:projection}). Moreover, it is the most straightforward method that is both local and directly conserves ions. Similarly, when removing a solid site, ions in the newly created fluid are simply set to zero.

It is important to notice that as the colloid moves, the number of solid sites it occupies is not constant; in order to keep a constant and homogeneous charge on the colloid, we redistribute $Q^i$ at every time-step so that the colloidal charge at each site $q_c(\vect{r}_i) =  (Q^i(\vect{r}_i) / v_p) \psi_s (\vect{r}_i)$, with the particle volume $v_p = (3/4) \pi r_p^3$.

Finally, the permittivity of the particles $\varepsilon^p(\vect{r})$ also has to be considered when solving Poisson's equation. For sites that are partially covered by the particles we take a simple interpolation using the solid fraction field between the permittivity of the fluids (already an interpolation depending on fluid concentration, as specified in Eq.~\eqref{eq:dielectric_constant}), and the permittivity of the particles, such that
\begin{equation}
	\varepsilon(\vect{r}_i) = \varepsilon^p(\vect{r}_i) \psi_s(\vect{r}_i) + \varepsilon^f(\vect{r}_i) (1-\psi_s(\vect{r}_i)).
\end{equation}

\section{Test cases} \label{sec:results}

\subsection{Fluid-fluid interface}

In this section the ion distribution next to a planar fluid interface is investigated. We validate the ion diffusive flux and corresponding hydrodynamic force given by solvation. Two different solvents are considered, which have different Gibbs transfer energies $\Delta \mu^\pm \ne 0$, as well as different permittivities, $\varepsilon^a \ne \varepsilon^b$. The results are compared with an analytical solution for ion distributions at interfaces derived by Onuki \textit{et al}~\cite{onuki2006ginzburg}. Bier et al. give a solution for the electrostatic potential of an interface at $x=0$, with fluid $a$ for $x < 0$ and $b$ for $x>0$~\cite{bier:2008}:
\begin{multline}
	\label{eq:pbpotential}
	\phi_\text{PB}(x) = \\
	\begin{cases}
		\phi_D - \tfrac{\phi_D}{A} e^{\kappa_a x},&x<0, \\
		\phi_D - \tfrac{\phi_D}{A} \left[\cosh(\kappa_i x)+b\sinh(\kappa_i x)\right],&x\in\left[0,s\right], \\
		\tfrac{\phi_D}{A} e^{-\kappa_b(x-s)} p \left[b \cosh(\kappa_i s)+\sinh(\kappa_i s)\right],&x>s.
	\end{cases}
\end{multline}
Here $A = (1+b\epsilon)\cosh(\kappa_i s) + (b+\epsilon)\sinh(\kappa_i s)$; $b = \sqrt{\varepsilon^a/\varepsilon^b}$; $\epsilon=\sqrt{n_a/n_b}$, with the subscript denoting the bulk values of the respective solvent. The screening lengths are given by $\kappa_{a,b} = (2\beta e^2n_{a,b}/\varepsilon^{a,b})^{1/2}$ and at the interface $\kappa_i = \sqrt{2 \beta e^2n_b/\varepsilon^a}$. The potential difference between the two bulk phases $\phi_D \equiv \phi_b - \phi_a$ is the Donnan potential. The parameter $s$ shifts the potential as a way to capture interfacial effects~\cite{bier:2008}, such as a smoothly varying $\varepsilon(x)$, present in our case.

Here and in all subsequent systems we take the same relaxation time for both fluids, $\tau^\sigma = 1.0 \Delta t$. Initial salt concentrations are given by $n_b/n_a=\exp(-\beta\Delta\mu_{\text{av}})$ with $\Delta\mu_\text{av} = (\Delta \mu^+ + \Delta\mu^-)/2$. The system is taken to be quasi-one-dimensional, $\{l_x,l_y,l_z\} = \{500\Delta x,4\Delta x,4\Delta x\}$, and periodic, so that there are two identical interfaces at $l_x = 0$ and $l_x = 250\Delta x$. The ion distributions are given by $n^\pm_\text{PB}(x) = n_a \exp[\mp \beta e \phi_\text{PB}(x) -\beta \mu^\pm_\text{s}(x)]$~\cite{onuki2006ginzburg}.

\begin{figure}
	\begin{center}
		\includegraphics[scale=1.0]{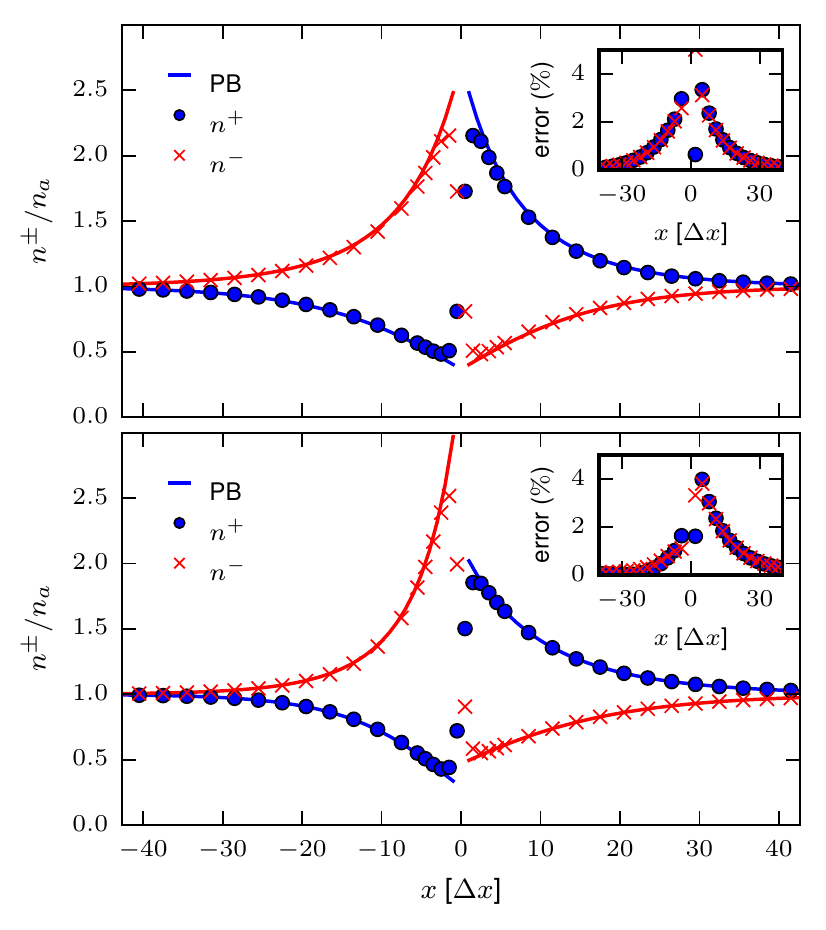}
	\end{center}
	\caption{Ionic density profiles $n^\pm(x)$ at a planar interface for $\Delta \mu^\pm = \pm 2 k_B T$, $\xi = 0$ (top) and $\Delta \mu = \pm 2 k_B T$, $\xi = 0.5$ (bottom). Insets show the relative error between the analytic and numerical solution, $|n^\pm - n^\pm_\text{PB}|/n_\text{PB}^\pm$.}
	\label{fig:interface01}
\end{figure}

Our simulations recover the predicted ionic concentrations to within $2\%$, except next to the interface sites $x \in [-5,5]\Delta x$, where the theoretical approximation of a sharp interface breaks (see Fig.~\ref{fig:interface01}). The error increases linearly in the $\Delta \mu^\pm \in (1,5)k_BT$ range, and stays below $5\%$ also for cases with different permittivity, when the permittivity constrast $\chi \equiv (\varepsilon_a - \varepsilon_b)/(\varepsilon_a+\varepsilon_b) \in (0,0.9)$.

The same setup was previously considered by Capuani \textit{et al}~\cite{rotenberg:2010}, where solvent/solvent interactions were modeled using a free-energy based on the concentration field $c(\vect{r})$, instead of using Shan-Chen pseudopotentials as in this study. In this respect, the most significant difference between the two approaches is the equilibrium state of a demixed mixture. While in our case there is a small but significant fraction of the minority fluid at the bulk phases, the concentration field in the free-energy formalism is truncated such that at the bulk it takes the perfectly demixed values, $|c| = 1$~\cite{rotenberg:2010}. Nevertheless, the theoretical derivation of Onuki does not assume perfectly demixed phases~\cite{onuki2006ginzburg}. In our case, agreement with the theoretical predictions was found only for the solvation potential normalized by the differences in concentration $\Delta \rho^a$, as in Eq.~\eqref{eq:solvation_potential}. When the expression in Ref.~\cite{rotenberg:2010} is used, which assumes perfect species separation, the convergent values of $\phi_\text{PB}(x=\pm\infty$) are not recovered, as the Donnan potential in that case does not take the form $(\Delta \mu^- - \Delta \mu^+)/2e$. We thus remark that the solvation potential in Eq.~\eqref{eq:solvation_potential} is a more general case that does not assume perfect phase separation.

\subsection{Dielectric droplet deformation}
\label{sec:dielectric_droplet}

In the following test, a droplet of solvent $b$ (oil) immersed in solvent $a$ (water) is considered in an external electric field $\vect{E}_\text{e} = (0, 0, E_z)$. In the absence of an electric field, the balance of surface tension and pressure determines the shape of the droplet, which minimizes the surface area: a sphere in 3D, a disk in 2D. When an external field is applied and the permittivities of the solvents are different, the droplet can elongate either in the direction of the applied field or perpendicular to it, depending on the permittivity contrast. The final shape is set by a balance between Laplace's pressure and the Kelvin force density. Notice that, as the system is ion-free, it allows us to independently test the dielectric force applied to the solvents, as in \eqref{eq:electric_fluid_force} the only non-vanishing term is the Kelvin force $\vect{F}_{E}$, Eq.~\eqref{eq:kelvin}.

As deformations are expected to be symmetric in the axis of the field, as also to optimize the computation time, only quasi-two-dimensional geometries are considered, by setting $l_x = 4 \Delta x$, while $l_y = l_z = l$. The deformation $\delta$ is defined as the normalized difference of half the droplet length in the $\hat y$ and $\hat z$ direction, $\delta \equiv (b-a)/(b+a)$, with $a$ in $\hat y$ and $b$ in $\hat z$. For small deformations the equilibrium shape is an ellipse, and thus $a$ and $b$ correspond to the semi-minor and semi-major axes, respectively. We compare our results to the existing analytical solution in the limit of small deformations~\cite{okonski:1953},
\begin{equation}
	\label{eq:droplet_deformation}
	\delta_\text{th} = \frac{\varepsilon^a r_d \chi^2 E_z^2}{4 \gamma},
\end{equation}
with $r_d$ the undeformed droplet radius and $\gamma$ the surface tension. The surface tension is obtained using the Young-Laplace equation, $\Delta p = \gamma/r_d$, where the pressure difference $\Delta p$ is taken to be between the center of the droplet and a point outside, computed from the equation of state of the Shan-Chen model, Eq.~\eqref{eq:eqofstate}. The size of the droplet and thus $\delta$ is determined at sub-lattice resolution by fitting the local concentration profile $c(y,z)$ next to the droplet interface by a hyperbolic tangent, and defining the interface position as the points where $c = 0$.

As in this configuration all electric forces, except for dielectric, are zero, the degree of the deformation is determined by the relative magnitude of the electric forces to the interfacial tension stresses, characterized by the electrocapillary number $\text{Ca}_{E} = \varepsilon^a r_d E_z^2 / \gamma$. The predicted deformation can thus be rewritten as $\delta_\text{th} = \text{Ca}_{E} \chi^2 / 4$.

\begin{figure}
	\begin{center}
		\includegraphics[scale=1.0]{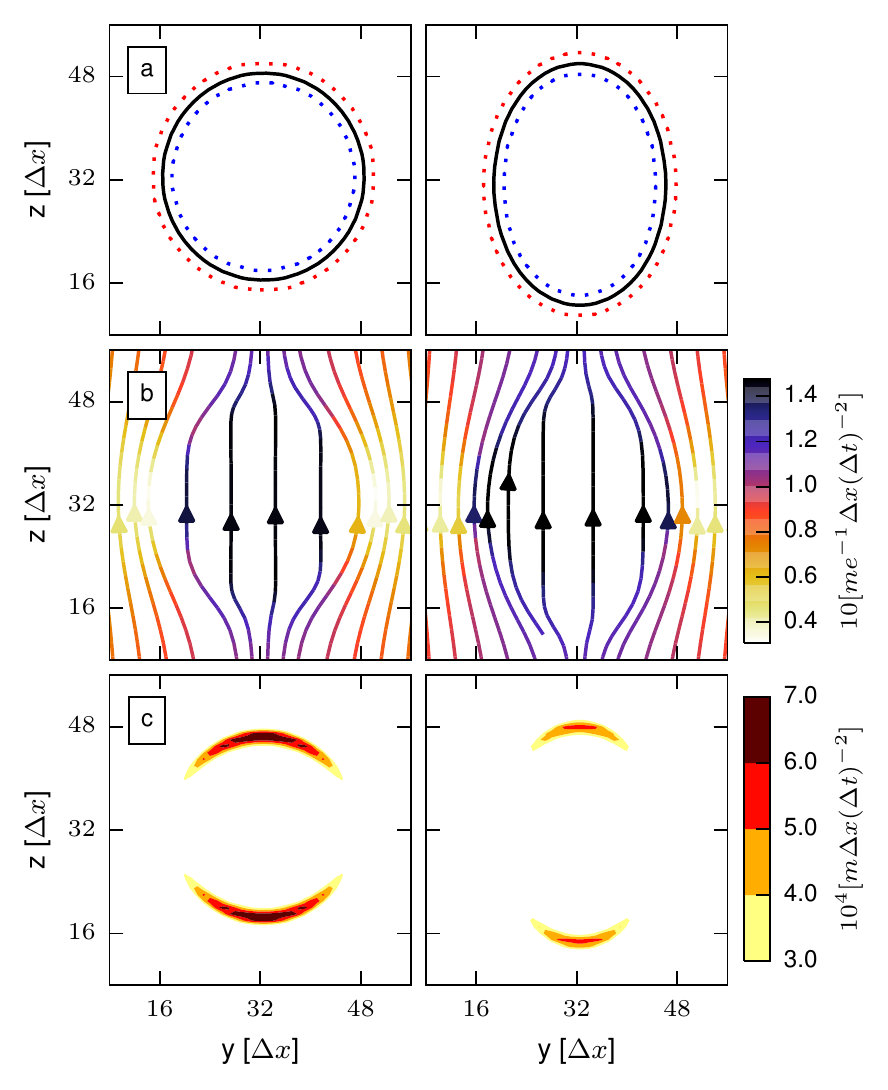}
	\end{center}
	\caption{(a) Contour lines of the concentration field $c(y,z) = 0.0$ (solid) and $c(y,z) = \{-0.8,0.8\}$ (dotted), for $\tilde E_z = 0.1$. (b) Stream lines of $\vect{E}(\vect{r})$. (c) The norm of the electric force $10^4|\vect{F}_{e}|$. All fields are presented at different times, $t = 1\Delta t$ (left), and in equilibrium at $t = 10^6 \Delta t$ (right), for $\chi = 0.9$.}
	\label{fig:drops}
\end{figure}
 
Boundary conditions are taken to be periodic, and thus the simulation can be considered to be of a square lattice of identical droplets. We study the system size dependency by considering two normalized droplet sizes, $\tilde r_d \equiv r_d/l \in \{0.1, 0.25\}$, for each system size $l \in \{64\Delta x, 128\Delta x\}$, corresponding to volume fractions $\nu \in (0.03, 0.2)$. Initially a circular oil region is set to relax for $10^5\Delta t$, and each simulation is then run from this equilibrated state for $2\times10^5\Delta t$. Both the strength of the applied field $E_z$ and the permittivity contrast $\chi$ are varied.

The droplet is seen to deform in the direction of the applied electric field, as shown in Fig.~\ref{fig:drops}a. The effect of the differences in permittivity is evident when looking at the electric field stream lines (Fig.~\ref{fig:drops}b), which result in a dominant dielectrophoretic force at the poles of the drop in the $\hat z$ direction (Fig.~\ref{fig:drops}c). This force is eventually balanced by the increase in surface tension due to the higher curvature of the deformed droplet.

\begin{figure}
	\begin{center}
		\includegraphics[scale=1.0]{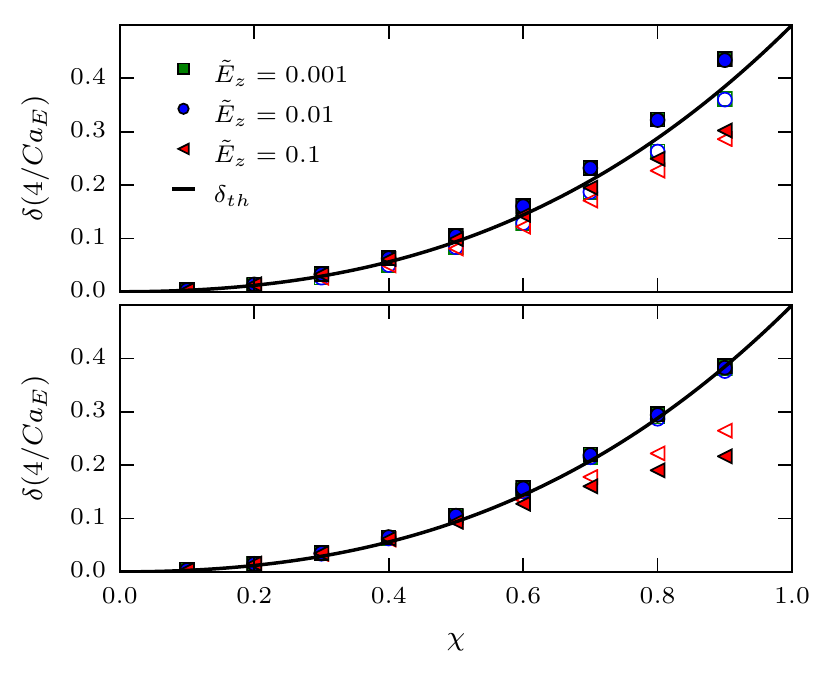}
	\end{center}
	\caption{Deformation $\delta$ of a droplet for different permittivity contrasts $\chi$ and dimensionless external electric field strengths $\tilde E_z$, normalized by the electrocapillary number $\text{Ca}_{E}$. The size of the square container is $l = 64 \Delta x$ (filled symbols) and $l = 128 \Delta x$ (empty symbols), for droplet size $\tilde r_d = 0.1$ (top) and $\tilde r_d = 0.25$ (bottom). The solid line shows the analytic solution, Eq.\eqref{eq:droplet_deformation}.}
	\label{fig:deltagamma}
\end{figure}

An excellent agreement between the computed deformation and $\delta_\text{th}$ is obtained for $\tilde E_z < 0.1$ and $r_d \ge 30 \Delta x$ (see Fig. \ref{fig:deltagamma}). (Here we have defined, for clarity, the dimensionless electric field $\tilde E_z = E/E_0$, with $E_0 = m_0e^{-1}\Delta x \Delta t^{-2}$). In these ranges, the normalized deformation is seen to be independent of $\tilde E_z$ and shows the expected quadratic dependence with $\chi$. This dependence breaks for $\tilde E_z > 0.1$ and high $\chi$; we do not expect Eq.\eqref{eq:droplet_deformation} to be valid for these deformations. Smaller droplets show errors within $15\%$ even for small $\tilde E_z$, most probably as a consequence of the finite-size diffuse interface. The results are consistent with those obtained in~\cite{rotenberg:2010}, where the solvent's interactions are modeled through a free-energy model, in such a way that the interface width and the surface tensions are parameters of the system. In our case, both of these quantities are a result of the Shan-Chen model, solely determined by the interaction parameter $G$, and are measured from the resulting equilibrium state.

\begin{figure}
	\begin{center}
		\includegraphics[scale=1.0]{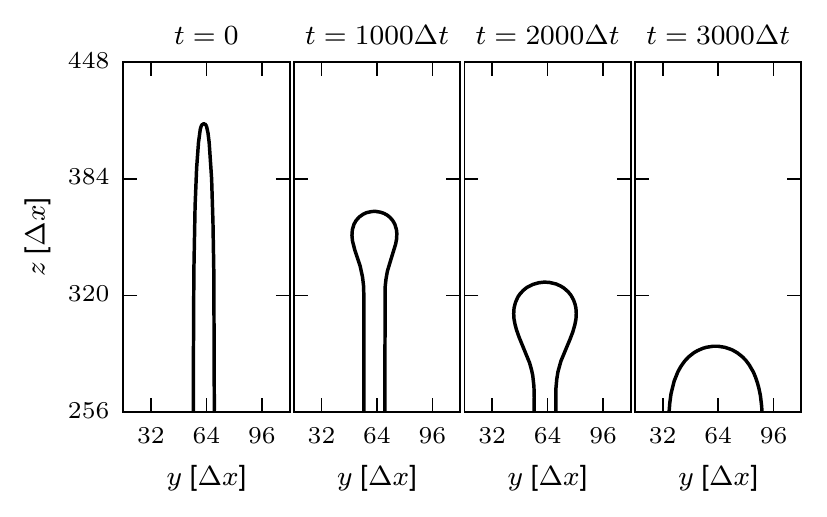}
	\end{center}
	\caption{Contour line of the concentration field $c(y,z) = 0.0$, of a deformed drop after the electric field is set to zero. Initially $\tilde E_z = 1.5$, and $\chi = 0.9$. As deformations are symmetric, only half of the system is shown.}
	\label{fig:drop_release}
\end{figure}

Increasing the electric field further leads to large elongated droplets with pointed ends, as shown in Fig.~\ref{fig:drop_release}. Similar shapes have been observed in experiments for drops under shear for low viscosity ratios between the two fluids~\cite{stone:1994}. When the field is turned off, the drops return to their equilibrium shape, and no breakup is observed even for the highest $\tilde E_z$ considered. 

\subsection{Neutral electrolyte droplets}

In the previous case no ions were considered, the deformation of the droplet being driven by the difference in permittivity between the two fluids. In the following, ions are added to the fluids, and the deformation of a drop in an electric field is studied. Modifying the Gibbs transfer energies $\Delta \mu^\pm$ allows us to produce ratios of bulk ion concentrations between the two fluids of up to $100$. At this limit we expect the influence of the ions in the outer fluid to be negligible, and therefore to be close to the most common experimental configuration where the drop's fluid has a much higher conductivity than the medium. In order to isolate the effects of the ions from other contributions, we take both fluids to have equal permittivities, $\chi = 0$, as well as equal densities and relaxation times, implying equal viscosities. We investigate the qualitative aspects of the deformation for variations in the average concentration $\bar n = \int n(\vect{r}) d\vect{r} / l_x l_y l_z$, and $E_z$. 

As the total electric force applied on the droplet now has three contributions (cf. Eq.~\eqref{eq:electric_fluid_force}), it is not evident that $\text{Ca}_{E}$ retains its relevance as it, for example, ignores the effects of increasing ion concentrations, which increase the electrostatic forces that lead to deformation.

We perform simulations of quasi-two-dimensional periodic domains of size $\{l_x, l_y, l_z\} = \{4\Delta x,128\Delta x,384\Delta x\}$. The system is initialized with a circular droplet of fluid $b$ of size $r_d = 32\Delta x$, and a homogeneous ion concentration $\bar n = 10^{-3} \Delta x^{-3}$. We first perform a relaxation phase during $2\times10^5\Delta t$ with $E_z = 0$, until diffusive fluxes vanish and the ion densities reach an equilibrium distribution. The Gibbs transfer energies are set to $\Delta \mu^\pm = -4 k_B T$, which results in a ratio of bulk ionic densities between the two species of $n_b/n_a \sim 80$. For simplicity we set equal permittivities on both fluids, such that $\varepsilon_r^f = 80$, with $\varepsilon^f_{r} = \varepsilon^f/\varepsilon_0$. Comparing the measured Debye length $\lambda_D = \sqrt{\varepsilon^f k_B T/e^2 \bar n}$ with experimental values known for water gives $\Delta x = 1\text{nm}$, for ionic concentrations in the dilute limit $10^{-3}\text{mol/l}$.

Although we expect finite-size effects to be relevant, we observe that the total electric field $\vect{E}$ at the boundaries is much smaller than $E_z$, and thus the qualitative aspects are not expected to be significantly altered by the domain size. 

\begin{figure}
	\begin{center}
		\includegraphics[scale=1.0]{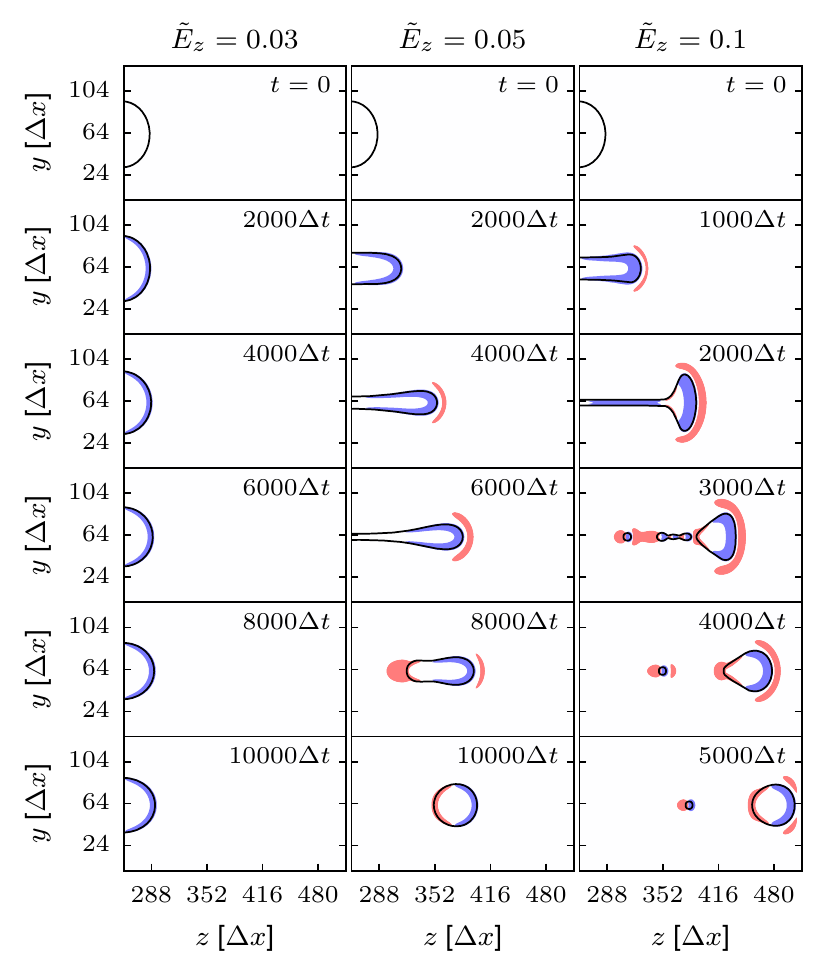}
	\end{center}
	\caption{Contour line $c(\vect{r}) = 0.0$ (black), for electrolyte droplets being deformed under the action of external electric fields of different strength. Also shown are contour regions for $q(\vect{r}) = -10^{-3}e$ (red) and $q(\vect{r}) = 10^{-3}e$ (blue). As deformations are symmetric, only half of the system is shown.}
	\label{fig:neutral_droplets}
\end{figure}

The electric field polarizes the drop as ions with different charges move in opposite directions. As a consequence, the electrostatic force, acting on opposite directions at each pole, deform the droplet, as shown in Fig.~\ref{fig:neutral_droplets}. For low $E_z$ the drop elongates and reaches stable ellipsoidal shapes. Beyond a critical $E_z$ the deformed drop becomes unstable, and after acquiring a dumbbell shape it breaks up in a pinch-off manner~\cite{nganguia:2016}. The morphology of the breakage varies with $E_z$:  slow elongations lead to round drop ends and thick connecting necks, with no satellite drops formed at breakup. Higher velocities, for higher applied fields, lead to greatly deformed ends perpendicular to the flow direction, a consequence of strong fluid drag and dielectric forces pushing in opposite directions. Capillary forces eventually break the slender neck, and two or more satellite drops are formed. Overall the qualitative aspects of the deformation and breakup coincide with previous experimental observations~\cite{ha:1998}, and numerical solutions of leaky-dielectric models in various limits~\cite{lac:2007,supeene:2008,nganguia:2015,nganguia:2016}. Here we have shown the capability of the model to study drop breakup in electric fields; further studies could quantify the effects of ion distributions at different concentration limits, providing information unreachable by previous numerical methods which assume all charges to be at the interface~\cite{pillai:2016}. Future quantitative studies should also take into account the known dependency of the degree of deformation and breakup with the grid resolution~\cite{komrakova2014lattice}.

\subsection{Charged droplets}
\label{sec:charged_droplets}

We now look at the deformation of droplets with a non-zero electric charge. A quasi-two-dimensional periodic box of size $\{l_x, l_y, l_z\} = \{4\Delta x, 384\Delta x, 384\Delta x\}$ with a droplet of radius $r_d = 32\Delta x$ is considered. The same relaxation procedure as in the previous system leads to the droplet having a negative net charge, when setting different Gibbs transfer energies for the two ion species $\Delta \mu^\pm = \pm 4 k_B T$. The concentration of ions is fixed at $\bar n = 10^{-3} \Delta x^{-3}$, which sets the average charge in the droplet at $\bar q_d  \sim - 3\times10^{-3}e$. The permittivities are set as in the previous systems, $\varepsilon_r = 80$. We vary the electric field $E_z$.

\begin{figure}
	\begin{center}
		\includegraphics[scale=1.0]{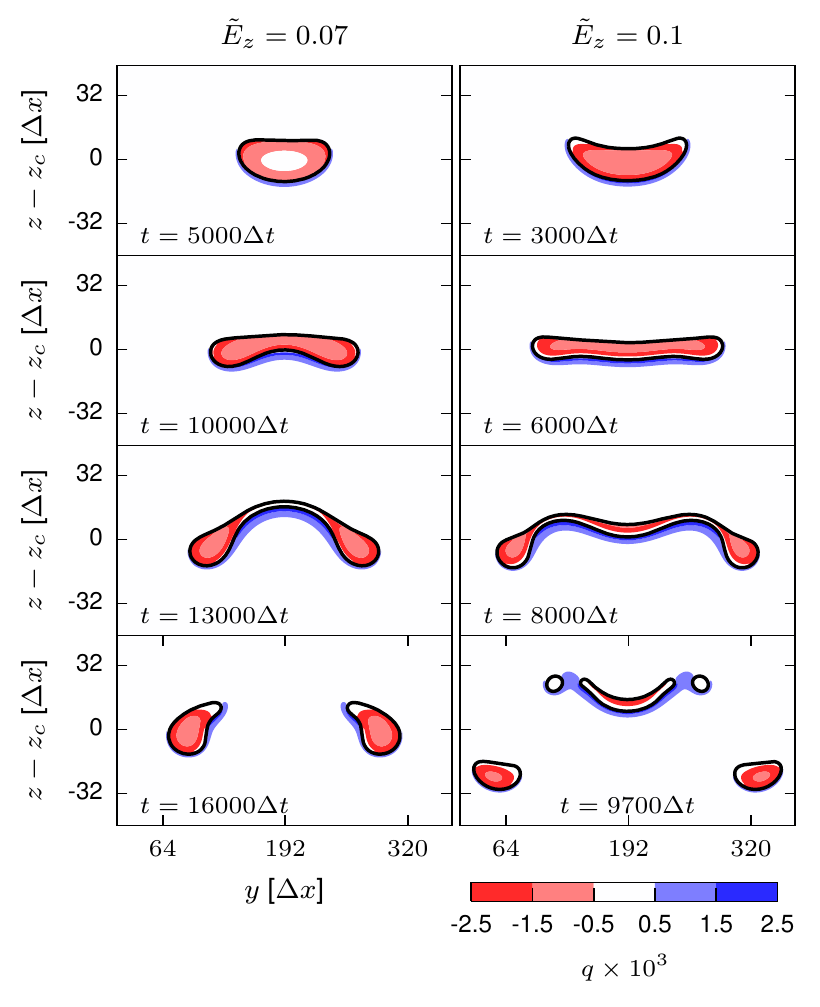}
	\end{center}
	\caption{Contour line $c(\vect{r}) = 0.0$ (black), for electrolyte droplets being deformed under the action of an external electric field $\tilde E_z = 0.07$ (a) and $\tilde E_z = 0.1$ (b). Also shown are contour surfaces of the total charge $q(\vect{r})$. To simplify the visualization, the $z$-direction is shown relative to the center of mass of $\rho^b$, $z_c$.}
	\label{fig:charged_droplets}
\end{figure}

As expected, drops accelerate in the direction of the electrostatic force $q \vect{E}$. Eventually they either reach a terminal velocity with an equilibrium deformed shape, or go through a breakup process. Initially the spherical shape widens perpendicular to the direction of flow, acquiring an oblate spheroidal shape~\cite{kulkarni:2014}. Afterwards, the drop bends, acquiring a bell-like shape with lobules at the ends connected by an increasingly long and thin neck, as shown in Fig.~\ref{fig:charged_droplets}. The ultimate breaking process is analogous to pinch-off, when surface tension forces break the thin neck. Depending on the strength of the field, at breakup the neck either withdraws to the end lobules, in which case only two drops are produced (Fig.~\ref{fig:charged_droplets}a), or breaks up further into smaller drops, a situation seen for the highest $E_z$ considered (Fig.~\ref{fig:charged_droplets}b).

Deformation and breakup are a result of hydrodynamic forces due to the relative velocity of the drop with the ambient fluid~\cite{guildenbecher:2009}. Previous studies of macroscopic drops and/or bubbles often refer to this process as secondary atomization, secondary breakup or droplet disintegration~\cite{gelfand:1996,guildenbecher:2009}. We observe that the breakup process for a charged droplet in an electric field is similar to the bag-breakup process known from previous experiments and numerical simulations of drops or bubbles flowing in either a fluid or a gaseous medium. After initial deformation, the center of the drop gets pushed downstream forming, in three-dimensions, a bag attached to a toroid. In our two-dimensional case, the analogous situation is of the observed neck connecting two separate lobules. Eventually the bag breaks, and is observed to fragment into a number of smaller drops~\cite{gelfand:1996,guildenbecher:2009}. We observe fragmentation only for the highest electric fields (see Fig.~\ref{fig:charged_droplets}), most probably as a consequence of the resolution limiting the size of smaller drops. In this last case, the bag is observed to create a plume at its center, most probably a precursor of a breakup process referred to as multimode breakup in experiments which, consistently, is seen to occur when increasing the forcing beyond bag breakup~\cite{guildenbecher:2009}.

The similarities of charged and uncharged phenomenologies can be further supported by considering the observed invariance of the breakup process with two dimensionless numbers: the Weber number, defined as the ratio of inertia to surface tension, $\text{We} = \rho^a u_d^2 r_d/\gamma$, with $u_d$ the terminal velocity of the drop; and the Ohnesorge number, taken as the ratio of the drop's viscous to surface tension forces, $\text{Oh} = \eta^a/\sqrt{\rho^a r_d \gamma}$. In our systems, $\text{Oh} \in (0.1,1.0)$, and $\textit{We} \in (10,30)$. At these values, various experimental studies have found the breakup process to be at either the bag or multimode type~\cite{gelfand:1996,guildenbecher:2009}.

\begin{figure*}
	\begin{center}
	    \includegraphics[scale=1.0]{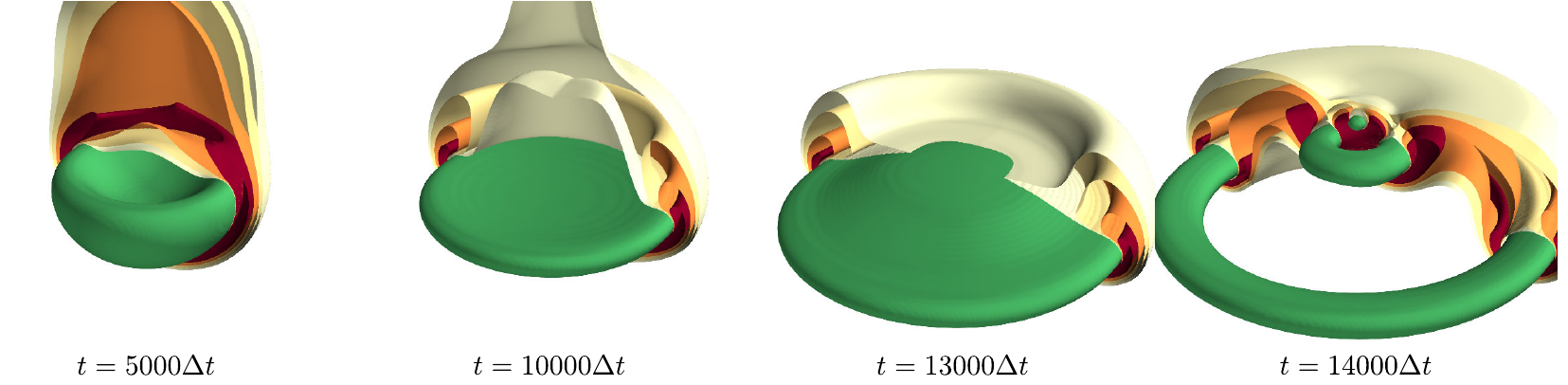}
	\end{center}
	\caption{Snapshots of a charged drop being deformed in an electric field, showing the drop interface, $c(\vect{r}) = 0.0$ (green), and contours for the charge concentration $q(\vect{r}) = \{1e,2e,3e,4e\}\times10^{4}$ (yellow to red). Here $r_d = 64\Delta x$, $\Delta \mu^\pm = \pm4.0k_B T$, $\bar n = 10^{-3} \Delta x^{-3}$ and $\tilde E_z = 0.07$.}
	\label{fig:charged_droplets3d}
\end{figure*}

Until now only quasi-two-dimensional systems have been considered, as the deforming droplets were always expected to be axisymmetric in the direction of the applied field. We wish to remark that our numerical implementation is performant enough to simulate three-dimensional systems of the cases previously presented. We realized individual simulations of selected cases and always observed qualitatively equivalent behaviours. As an example, we show in Fig.~\ref{fig:charged_droplets3d} the break-up process of a charged droplet of size $r_d = 64\Delta x$, with parameters as in Fig.~\ref{fig:charged_droplets}b, in a periodic cube of side length $l = 384\Delta x$. In this geometry one can clearly see how the elongated-disk drop breaks initially at its center, forming a toroid. Negative charges, initially forming a spherical double layer, are quickly accumulated at the front of the droplet. The fluid flux also significantly advects the charges through the outer and inner (after breakup) sides of the drop. Overall, the breakup process is consistent with the quasi-two-dimensional drops.

In summary, we conclude that the effect of the double layer on the dynamics of breakup is not highly significant in this parameter range, as the drop breakup process is to a large extent analogous to a non-charged drop deformed under the effect of  hydrodynamic forces. Nevertheless, we do not expect this to be the case for other salt concentration limits. In general we have shown that our numerical methodology is capable of capturing the qualitative aspects of this process in the right order of the relevant dimensionless numbers. Future studies could systematically vary the concentration of ions to quantify the influence of electric forces on drop deformation and breakup.

\subsection{Colloid electrophoresis}

In the following sections electrokinetic colloidal suspensions are studied. In order to validate the coupling of the colloidal particles to solute and solvents, we first consider the basic electrophoretic setup in which a single, charged particle moves in an electrolyte solution under the effect of an electric field. We compare our results with experimental measurements~\cite{lobaskin:2007} and lattice Boltzmann simulations~\cite{giupponi:2011}. In the latter study~\cite{giupponi:2011}, the link-flux method was used to solve the electrokinetic equations in a similar manner as in the present study although, importantly, only fixed particles were considered, while in our case the particles are free to move.

The velocity at which a colloidal particle with charge $Q$ moves in a weak electric field is linear with $E$, with the proportionality constant refered to as the electrophoretic mobility $\mu = v/E$. A colloidal particle is usually surrounded by counter-ions (ions with a total charge $-Q$), which create an electric field opposing the movement of the particle. The charges on the colloid and the cloud of counter-ions is referred to as double layer. The rest of the electrolyte ions in the solution, referred to as salt or co-ions, increase the drag on the particle and decrease the electric field, thus reducing the mobility $\mu$ of the particle relative to the case with no salt~\cite{obrien:1978}. The theoretical derivation of $\mu$ as a function of the charge of the particle $Q$, the salt concentration and, in the case of solutions, the solid fraction $\eta$ is not trivial, and has been done only for the limiting cases of small or large double layer thickness~\cite{obrien:1978}. In the following we validate our simulations by comparing $\mu$ with previous numerical work and experiments, usually realized at the mili- or microscale.

A single colloid of radius $r_p = 4.0 \Delta x$ is situated in a periodic box of size $l$. Due to the use of periodic boundary conditions the system can be considered as a cubic colloidal crystal of solid fraction $\eta = (4/3) \pi r_p^3/l^3$. We modify the solid fraction by varying $l \in (24,128)\Delta x$, that is $\nu \in (10^{-4},10^{-2})$, and set the material parameters as in previous experiments and simulations~\cite{lobaskin:2007,giupponi:2011}: colloid charge $Q = 30e$, and a Bjerrum length $\lambda_B = e^2/(4\pi \varepsilon k_B T) = 1.3 \Delta x$. The permittivity of the fluids and particles is set such that $\varepsilon^f_{r} = 80$, and $\varepsilon^p_{r} = 10$.

\begin{figure}
	\begin{center}
		\includegraphics[scale=1.0]{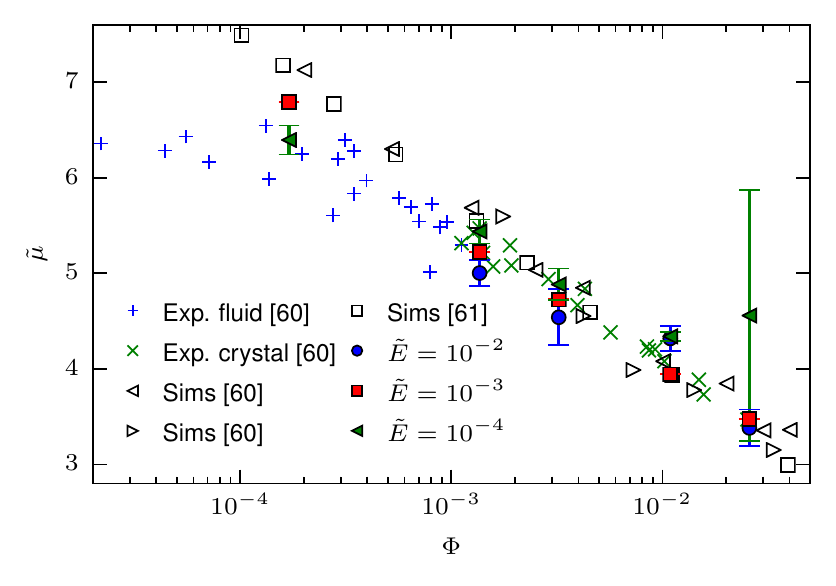}
	\end{center}
	\caption{Dimensionless electrophoretic mobility $\tilde \mu$ as a function of the solid fraction $\nu$, for a colloidal particle of size $r_p = 4.0 \Delta x$, and charge $Q = 30e$. Also shown is experimental data for suspensions of latex particles of diameter $64\text{nm}$ in a fluid ($+$) or crystalline ($\times$) state, as detailed in~\cite{lobaskin:2007}. Furthermore, data is shown for simulations using a lattice-Boltzmann/Molecular Dynamics combination as described in \cite{lobaskin:2007}, for particles of total charge $Q = 20e$ and radius $4\Delta x$ ($\triangleleft$), and $Q = 30e$ and radius $6 \Delta x$ ($\triangleright$). Finally, data for simulations using an equivalent algorithm as this study, although with fixed particles, is also included ($\square$)~\cite{giupponi:2011}.}
	\label{fig:muphi}
\end{figure}

A good agreement is observed with the previously measured dependency of the electrophoretic mobility with the solid fraction, $\mu(\nu)$, as shown in Fig.~\ref{fig:muphi}. As in previous works, we have expressed our results using the dimensionless mobility $\tilde \mu = 6 \pi \eta l_B \mu / e$. Consistent deviations with experimental values increase with the applied field, although they remain within a $5\%$ error for $\tilde E_z \in (10^{-4}, 10^{-2})$. We expect the deviations at high forcing to be a consequence of the fluid terminal velocity being close to $c_s$, namely $\text{Ma} \sim 1$, where many of the model assumptions are no longer valid. On the other hand, for very low fields, $\tilde E_z = 10^{-4}$, the very low velocity of the colloid is significantly affected by discretization effects, not only due to flux of ions but also due to the hydrodynamic boundary conditions. These generate large fluctuations in $\mu$ (as can be seen by the error bars in Fig.~\ref{fig:muphi}), although the average remains consistent with previous studies for $\nu > 10^{-2}$.

\subsection{A particle at a fluid interface}

As final exemplary cases for the possibilities of the numerical method, particles at interfaces of electrolyte solutions are studied. We begin by considering the behavior of a single particle at a planar fluid interface, and afterwards show the possibility of simulating colloidal suspensions on the surface of droplets. These cases represent common scenarios of scientific and technological relevance. Nanoparticles at fluid interfaces are common in various in-development technological advancements, such as advanced coating processes for molecular electronics and optical devices~\cite{caruso:2001,gurrappa:2008}, stabilization of emulsions~\cite{chevalier:2013} or better control and transport capabilities in micro- and nanofluidic devices~\cite{chang:2010}.

Colloidal particles at fluid interfaces are highly stable, as the particle-oil and particle-water surface tensions are in most cases much smaller than the oil-water one. Although the energy needed for detachment of the interface scales quadratically with the radius of the particle~\cite{pieranski1980two}, at the nanoscale the energies needed for detachment are still $10$-$100$ times larger than $k_B T$~\cite{lin2003nanoparticle,bresme2007nanoparticles}. Therefore we ignore, as a first approximation, temperature fluctuations.

A single particle of radius $r_p = 8 \Delta x$ is placed at a fluid interface in a periodic box of size $l = 64 \Delta x$. The radius of the particle was chosen to minimize the effects of the finite width of the interface. As the system is periodic, the setup is equivalent to an infinite crystalline colloidal suspension with solid fraction $\nu = 8 \times 10^{-3}$. The interface without the colloid is previously relaxed for $10^5\Delta t$, such that the concentration of ions, initially homogeneously set to $n(\vect{r}) = \bar n$, reaches an equilibrium distribution given by the different Gibbs transfer energies $\Delta \mu^\pm$. These are fixed at $\Delta \mu^\pm = 4.0 k_B T$. Permittivities are set as in the single colloid electrophoresis case, $\varepsilon^f_{r} = 80$, and $\varepsilon^p_{r} = 10$. We vary $\bar n$ and the charge of the colloid $Q$, and measure the displacement of the colloid from its equilibrium position with no solutes, $z^*_0$, $z^* = |z_c - z_0^*|$.

\begin{figure}
	\begin{center}
	\includegraphics[scale=1.0]{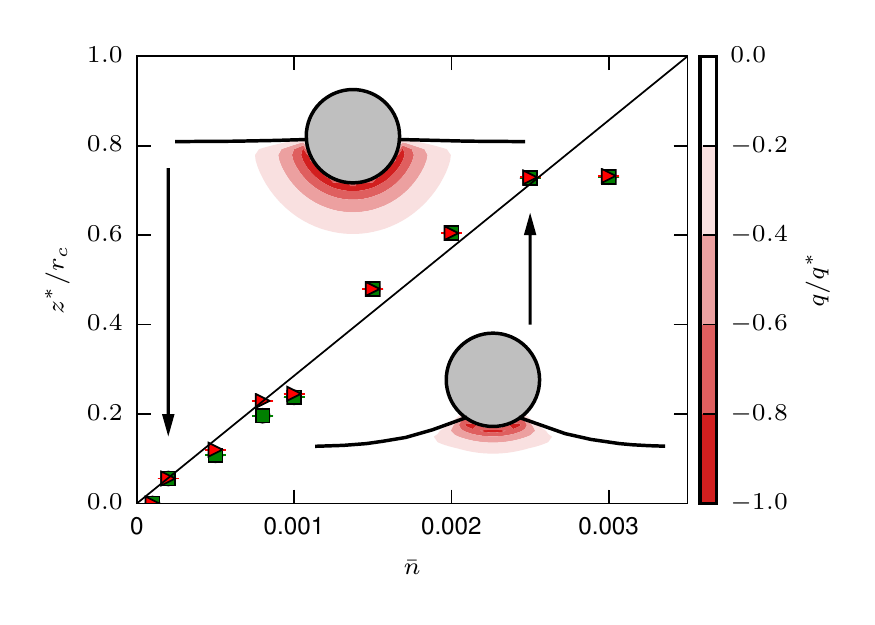}
	\end{center}
	\caption{Relative equilibrium position of a colloid at a fluid interface, normalized by its diameter, shown as a function of the salt concentration $\bar n$, for different colloid charges $Q=1e$ ($\square$) and $Q=10e$ ($\triangleright$). The solid line shows a linear dependency. The snapshots show the interface position $c(z) = 0$ as a solid line, the particle in gray, and the contours of concentration of charges, normalized the maximum charge in each case, for the limit cases indicated by the arrows and $Q = 10$.}
	\label{fig:eqzvssalt}
\end{figure}

The displacement of the particle is seen to increase linearly with $\bar n$, as shown in Fig.~\ref{fig:eqzvssalt}. At a critical salt concentration the ion's osmotic pressure is strong enough to detach the particle from the interface. Close to this limit, the linear behavior is lost, and the displacement saturates, probably an effect of the steep interface deformations interacting with the discretized particle shape. 

The equilibrium position of a particle at a fluid interface is usually quantified by the contact angle $\theta_c$. This is the angle formed by the tangent of the interface and the tangent to the particle at the triple contact point. We observe $\theta_c$ to be independent of the displacement of the particle. This is to be expected, as the wetting properties of the particle have not been modified, and even though the particle has displaced, the interface deforms in such a way that $\theta_c$ is kept constant, $\theta_c = 90^o$ in our case. In addition, we wish to remark that having access to $\theta_c$ in experimental situations at the nanoscale can present a significant challenge. More commonly the apparent contact angle $\theta_a$ is measured, formed by the tangent of the interface at the triple contact point with the horizontal. In that case we naturally observe a variation with salt concentration, as $\sin(\theta_a) = z^*/r_p$.

Surprisingly we see no strong effect of the charge of the particle in $z^*_0$, for all considered charges and salt concentrations. As shown in Fig.~\ref{fig:eqzvssalt}, charged particles generate a double layer, with most of the charge concentrated on the conducting side. Nevertheless, the resulting electrostatic force is observed to be at most an order of magnitude lower than both the ions osmotic pressure and the solvation forces. As the total concentration of ions around the particle is mostly independent of $Q$ (the increase in negative ions next to the colloid compensates the loss of positive ions), the total electric force $\vect{F}_e$ only slightly varies with $Q$, and therefore the equilibrium position is only marginally affected.

In summary, we have briefly shown the possibility of studying nanoparticle adsorbtion at fluid interfaces with resolved charge distributions, using the presented numerical methods. Future studies could explore the stability of the particle as a function of electrolyte ion concentration, particle charge, and the possibility of higher control using an external electric field. Research in this direction could help to improve our understanding of electrodipping forces at the nanoscale~\cite{nikolaides2002electric,danov2004electrodipping}, a necessary step to obtain a general interaction potential for particles at fluid interfaces.

\subsection{Colloid coated droplets}

Finally we demonstrate the possibility of simulating the dynamics of colloidal suspensions adsorbed at curved fluid interfaces, such as the surface of a droplet. Colloid-coated droplets are present in many physical and chemical processes. In Pickering emulsions, the coating of droplets by colloids can stabilize emulsions without the need of surfactants. Emulsions can then be used as templates for fabrication of capsules composed of colloids: colloidosomes~\cite{chevalier:2013,shah:2008,dommersnes:2013}. Colloidosomes can be functionalized as containers of different substances, such as drugs, general reagents, or even cells, with several potential applications in the food, biomedical and petroleum industries~\cite{dommersnes:2013}.

In this case we consider a droplet of radius $r_d = 32 \Delta x$ covered by an homogeneous suspension of particles of size $r_p = 5 \Delta x$. The droplet parameters are identical as for the deforming charged droplets studied Section~\ref{sec:dielectric_droplet}, as we wish to investigate the influence of the particles in the overall deformation of the droplet. For simplicity we consider all particles to be uncharged, $Q^j = 0$. The total number of particles is $N = 80$, which gives a covering fraction of $\nu_s = (1/4) N r_p^2 / r_d^2 \approx 0.48$. Permittivities are set as in the previous cases, $\varepsilon^f_{r} = 80$, and $\varepsilon^p_{r} = 10$.

\begin{figure}
	\begin{center}
	    \includegraphics[scale=1.0]{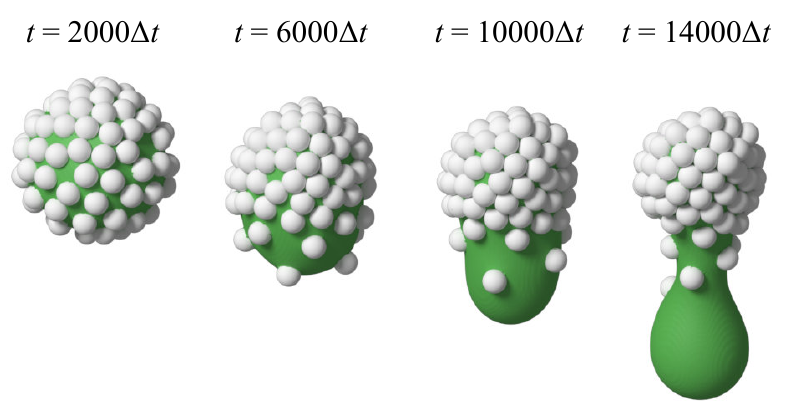}
	\end{center}
	\caption{Snapshots of a particle-coated charged droplet in an electric field. Parameters are identical as in Fig.~\ref{fig:charged_droplets3d}, with $r_p = 5.0 \Delta x$. The drop interface $c(\vect{r}) = 0.0$ (green) and colloids (white).}
	\label{fig:coatedrop}
\end{figure}

The effect of the covering particles on the deformation and breakup morphology of the droplet is dramatic. While without colloids the droplet underwent a process similar to bag breakup (see Fig.~\ref{fig:charged_droplets3d}), in this case the breakup process is more similar to pinch-off: the drop deforms and squeezes through the particle coating, forming a long neck until it eventually breaks, as shown in Fig.~\ref{fig:coatedrop}. The results is the creation of a new, particle-free droplet, and the increase of $\nu_s$ in the original droplet, such that it is now stable in the electric field. Including corrections to take into account the overlapping ion clouds at small distances would certainly reduce the equilibrium packing fractions, although we expect to observe the same qualitative dynamics. Due to the increased inertia of the coated droplets, their speed is considerably slower than the particle-free ones, so that drag forces do not significantly deform its shape. Overall we see phenomena worthy of future studies which might have practical significance, as the covering fraction of coated droplets could be increased by applying an external electric field and thus removing the excess fluid from the drop.

\section{Conclusions}
\label{sec:conclusions}

A mesoscopic model for the simulation of electrokinetic phenomena of colloidal suspensions in fluid mixtures at the nanoscale was presented. The model follows a more microscopic description than the link-flux model originally proposed by Capuani~\textit{et al}., treating the kinetics of the ions as a response to individual forces instead of using a free-energy functional, and deriving the coupling between the different species using local pseudopotentials. Overall we have shown that binary mixtures of electrolytes can be successfully modeled using the Shan-Chen multicomponent pseudopotentials for both fluid-fluid and fluid-ion interactions, providing a new, simple methodology for the simulation of such systems. Colloids are included via a Ladd coupling with the fluids, and a solid-fraction scheme of discretization that significantly reduces discretization errors. Results were shown to be consistent with the previous link-flux method, and agree with known theoretical solutions for ionic concentrations at fluid interfaces, droplet deformation and electrophoresis of a single colloid.

Several systems were shown to explore the posibilities of the model. Droplet deformation and breakup for neutral and charged droplets presents the same qualitative aspects as previous numerical and experimental studies at the microscale. The ability to integrate particles and fluid mixtures opens the possiblity to study at a new level of resolution the statics and dynamics of colloids at interfaces, both planar and curved. Our first investigations show that colloids at a planar electrolyte interface vary their equilibrium position when ions are added to one of the fluid, until the point were adsorbtion becomes unstable and the particle detaches. Furthermore, it is now possible to study (charged) colloidal suspensions adsorbed at the surface of a droplet; we have here shown a first exemplary case of a colloid-coated droplet breakup due to an external electric field. All these examples illustrate that the presented method can be used for studies of a variety of electrokinetic phenomena in parameter ranges so far unreachable.

This description of ion transport in electrolytes could be easily extended to include other effects. Additional forces acting on the ions can be readily added in Eq.~\eqref{eq:electric_ion_force}, with the ion-solvent coupling force following directly from the friction coupling. More involved extensions beyond the Poisson-Boltzmann limit of no ion-ion interactions or solvent polarization, could include the use of modified Poisson-Nernst-Planck equations (MPNP) that take steric effects into account~\cite{kilic:2007}. These could be included as additional terms in the ion flux, Eq.~\eqref{eq:ionicflux}, and are possibly crucial in confined systems at high ion concentrations or strong external fields~\cite{kilic:2007}.

In summary, our work has shown that mesoscopic simulations can be used to study electrokinetic phenomena at the nanoscale, where obtaining experimental data becomes a challenge.  The same algorithm could be applied to a variety of nanofluidic systems, such as droplet transport in inhomogeneous geometries, coalescence and generation of droplets, and nanomixers.

\section{Acknowledgements}

S. Frijters and J. Harting acknowledge financial support from the FOM/Shell IPP (09iPOG14 - ”Detection and guidance of nanoparticles for enhanced oil recovery”) and NWO/STW (Vidi grant 10787 of J. Harting). We thank the J\"ulich Supercomputing Centre and the High Performance Computing Centre Stuttgart for the technical support and the allocated CPU time. We further thank Peter Coveney, Gary Davies, Giovanni Giupponi, Oliver Henrich, Christian Holm, Florian Janoschek, Timm Kr\"uger, Michael Kuron, Gavin Pringle, Georg Rempfer, and Kevin Stratford for fruitful discussions and support with the software development.

\bibliography{numerics} 

\begin{thebibliography}{74}%
\makeatletter
\providecommand \@ifxundefined [1]{%
 \@ifx{#1\undefined}
}%
\providecommand \@ifnum [1]{%
 \ifnum #1\expandafter \@firstoftwo
 \else \expandafter \@secondoftwo
 \fi
}%
\providecommand \@ifx [1]{%
 \ifx #1\expandafter \@firstoftwo
 \else \expandafter \@secondoftwo
 \fi
}%
\providecommand \natexlab [1]{#1}%
\providecommand \enquote  [1]{``#1''}%
\providecommand \bibnamefont  [1]{#1}%
\providecommand \bibfnamefont [1]{#1}%
\providecommand \citenamefont [1]{#1}%
\providecommand \href@noop [0]{\@secondoftwo}%
\providecommand \href [0]{\begingroup \@sanitize@url \@href}%
\providecommand \@href[1]{\@@startlink{#1}\@@href}%
\providecommand \@@href[1]{\endgroup#1\@@endlink}%
\providecommand \@sanitize@url [0]{\catcode `\\12\catcode `\$12\catcode
  `\&12\catcode `\#12\catcode `\^12\catcode `\_12\catcode `\%12\relax}%
\providecommand \@@startlink[1]{}%
\providecommand \@@endlink[0]{}%
\providecommand \url  [0]{\begingroup\@sanitize@url \@url }%
\providecommand \@url [1]{\endgroup\@href {#1}{\urlprefix }}%
\providecommand \urlprefix  [0]{URL }%
\providecommand \Eprint [0]{\href }%
\providecommand \doibase [0]{http://dx.doi.org/}%
\providecommand \selectlanguage [0]{\@gobble}%
\providecommand \bibinfo  [0]{\@secondoftwo}%
\providecommand \bibfield  [0]{\@secondoftwo}%
\providecommand \translation [1]{[#1]}%
\providecommand \BibitemOpen [0]{}%
\providecommand \bibitemStop [0]{}%
\providecommand \bibitemNoStop [0]{.\EOS\space}%
\providecommand \EOS [0]{\spacefactor3000\relax}%
\providecommand \BibitemShut  [1]{\csname bibitem#1\endcsname}%
\let\auto@bib@innerbib\@empty
\bibitem [{\citenamefont {Li}(2004)}]{li:2004}%
  \BibitemOpen
  \bibfield  {author} {\bibinfo {author} {\bibfnamefont {D.}~\bibnamefont
  {Li}},\ }\href@noop {} {\emph {\bibinfo {title} {Electrokinetics in
  microfluidics}}},\ Vol.~\bibinfo {volume} {2}\ (\bibinfo  {publisher}
  {Academic Press},\ \bibinfo {year} {2004})\BibitemShut {NoStop}%
\bibitem [{\citenamefont {Wong}\ \emph {et~al.}(2004)\citenamefont {Wong},
  \citenamefont {Wang}, \citenamefont {Deval},\ and\ \citenamefont
  {Ho}}]{wong:2004}%
  \BibitemOpen
  \bibfield  {author} {\bibinfo {author} {\bibfnamefont {P.~K.}\ \bibnamefont
  {Wong}}, \bibinfo {author} {\bibfnamefont {T.-H.}\ \bibnamefont {Wang}},
  \bibinfo {author} {\bibfnamefont {J.~H.}\ \bibnamefont {Deval}}, \ and\
  \bibinfo {author} {\bibfnamefont {C.-M.}\ \bibnamefont {Ho}},\ }\href@noop {}
  {\bibfield  {journal} {\bibinfo  {journal} {IEEE/ASME transactions on
  mechatronics}\ }\textbf {\bibinfo {volume} {9}},\ \bibinfo {pages} {366}
  (\bibinfo {year} {2004})}\BibitemShut {NoStop}%
\bibitem [{\citenamefont {Viovy}(2000)}]{viovy:2000}%
  \BibitemOpen
  \bibfield  {author} {\bibinfo {author} {\bibfnamefont {J.-L.}\ \bibnamefont
  {Viovy}},\ }\href@noop {} {\bibfield  {journal} {\bibinfo  {journal} {Reviews
  of Modern Physics}\ }\textbf {\bibinfo {volume} {72}},\ \bibinfo {pages}
  {813} (\bibinfo {year} {2000})}\BibitemShut {NoStop}%
\bibitem [{\citenamefont {Siretanu}\ \emph {et~al.}(2014)\citenamefont
  {Siretanu}, \citenamefont {Ebeling}, \citenamefont {Andersson}, \citenamefont
  {Stipp}, \citenamefont {Philipse}, \citenamefont {Stuart}, \citenamefont {Van
  Den~Ende},\ and\ \citenamefont {Mugele}}]{siretanu:2014}%
  \BibitemOpen
  \bibfield  {author} {\bibinfo {author} {\bibfnamefont {I.}~\bibnamefont
  {Siretanu}}, \bibinfo {author} {\bibfnamefont {D.}~\bibnamefont {Ebeling}},
  \bibinfo {author} {\bibfnamefont {M.~P.}\ \bibnamefont {Andersson}}, \bibinfo
  {author} {\bibfnamefont {S.~S.}\ \bibnamefont {Stipp}}, \bibinfo {author}
  {\bibfnamefont {A.}~\bibnamefont {Philipse}}, \bibinfo {author}
  {\bibfnamefont {M.~C.}\ \bibnamefont {Stuart}}, \bibinfo {author}
  {\bibfnamefont {D.}~\bibnamefont {Van Den~Ende}}, \ and\ \bibinfo {author}
  {\bibfnamefont {F.}~\bibnamefont {Mugele}},\ }\href@noop {} {\bibfield
  {journal} {\bibinfo  {journal} {Scientific reports}\ }\textbf {\bibinfo
  {volume} {4}},\ \bibinfo {pages} {4956} (\bibinfo {year} {2014})}\BibitemShut
  {NoStop}%
\bibitem [{\citenamefont {Saville}(1997)}]{saville:1997}%
  \BibitemOpen
  \bibfield  {author} {\bibinfo {author} {\bibfnamefont {D.~A.}\ \bibnamefont
  {Saville}},\ }\href {\doibase 10.1146/annurev.fluid.29.1.27} {\bibfield
  {journal} {\bibinfo  {journal} {Annual Review of Fluid Mechanics}\ }\textbf
  {\bibinfo {volume} {29}},\ \bibinfo {pages} {27} (\bibinfo {year}
  {1997})}\BibitemShut {NoStop}%
\bibitem [{\citenamefont {Schnitzer}\ and\ \citenamefont
  {Yariv}(2015)}]{schnitzer2015taylor}%
  \BibitemOpen
  \bibfield  {author} {\bibinfo {author} {\bibfnamefont {O.}~\bibnamefont
  {Schnitzer}}\ and\ \bibinfo {author} {\bibfnamefont {E.}~\bibnamefont
  {Yariv}},\ }\href@noop {} {\bibfield  {journal} {\bibinfo  {journal} {Journal
  of Fluid Mechanics}\ }\textbf {\bibinfo {volume} {773}},\ \bibinfo {pages}
  {1} (\bibinfo {year} {2015})}\BibitemShut {NoStop}%
\bibitem [{\citenamefont {Liu}\ \emph {et~al.}(2012)\citenamefont {Liu},
  \citenamefont {Wang}, \citenamefont {Chen},\ and\ \citenamefont
  {Robbins}}]{liu:2012}%
  \BibitemOpen
  \bibfield  {author} {\bibinfo {author} {\bibfnamefont {J.}~\bibnamefont
  {Liu}}, \bibinfo {author} {\bibfnamefont {M.}~\bibnamefont {Wang}}, \bibinfo
  {author} {\bibfnamefont {S.}~\bibnamefont {Chen}}, \ and\ \bibinfo {author}
  {\bibfnamefont {M.~O.}\ \bibnamefont {Robbins}},\ }\href {\doibase
  10.1103/PhysRevLett.108.216101} {\bibfield  {journal} {\bibinfo  {journal}
  {Phys. Rev. Lett.}\ }\textbf {\bibinfo {volume} {108}},\ \bibinfo {pages}
  {216101} (\bibinfo {year} {2012})}\BibitemShut {NoStop}%
\bibitem [{\citenamefont {Chen}\ \emph {et~al.}(2013)\citenamefont {Chen},
  \citenamefont {Li}, \citenamefont {van~der Vegt}, \citenamefont
  {Auernhammer},\ and\ \citenamefont {Bonaccurso}}]{chen:2013}%
  \BibitemOpen
  \bibfield  {author} {\bibinfo {author} {\bibfnamefont {L.}~\bibnamefont
  {Chen}}, \bibinfo {author} {\bibfnamefont {C.}~\bibnamefont {Li}}, \bibinfo
  {author} {\bibfnamefont {N.~F.~A.}\ \bibnamefont {van~der Vegt}}, \bibinfo
  {author} {\bibfnamefont {G.~K.}\ \bibnamefont {Auernhammer}}, \ and\ \bibinfo
  {author} {\bibfnamefont {E.}~\bibnamefont {Bonaccurso}},\ }\href {\doibase
  10.1103/PhysRevLett.110.026103} {\bibfield  {journal} {\bibinfo  {journal}
  {Phys. Rev. Lett.}\ }\textbf {\bibinfo {volume} {110}},\ \bibinfo {pages}
  {026103} (\bibinfo {year} {2013})}\BibitemShut {NoStop}%
\bibitem [{\citenamefont {Malevanets}\ and\ \citenamefont
  {Kapral}(1999)}]{malevanets:1999}%
  \BibitemOpen
  \bibfield  {author} {\bibinfo {author} {\bibfnamefont {A.}~\bibnamefont
  {Malevanets}}\ and\ \bibinfo {author} {\bibfnamefont {R.}~\bibnamefont
  {Kapral}},\ }\href@noop {} {\bibfield  {journal} {\bibinfo  {journal} {The
  Journal of chemical physics}\ }\textbf {\bibinfo {volume} {110}},\ \bibinfo
  {pages} {8605} (\bibinfo {year} {1999})}\BibitemShut {NoStop}%
\bibitem [{\citenamefont {Allahyarov}\ and\ \citenamefont
  {Gompper}(2002)}]{allahyarov:2002}%
  \BibitemOpen
  \bibfield  {author} {\bibinfo {author} {\bibfnamefont {E.}~\bibnamefont
  {Allahyarov}}\ and\ \bibinfo {author} {\bibfnamefont {G.}~\bibnamefont
  {Gompper}},\ }\href@noop {} {\bibfield  {journal} {\bibinfo  {journal}
  {Physical Review E}\ }\textbf {\bibinfo {volume} {66}},\ \bibinfo {pages}
  {036702} (\bibinfo {year} {2002})}\BibitemShut {NoStop}%
\bibitem [{\citenamefont {Hecht}\ \emph {et~al.}(2005)\citenamefont {Hecht},
  \citenamefont {Harting}, \citenamefont {Ihle},\ and\ \citenamefont
  {Herrmann}}]{hecht:2005}%
  \BibitemOpen
  \bibfield  {author} {\bibinfo {author} {\bibfnamefont {M.}~\bibnamefont
  {Hecht}}, \bibinfo {author} {\bibfnamefont {J.}~\bibnamefont {Harting}},
  \bibinfo {author} {\bibfnamefont {T.}~\bibnamefont {Ihle}}, \ and\ \bibinfo
  {author} {\bibfnamefont {H.~J.}\ \bibnamefont {Herrmann}},\ }\href {\doibase
  10.1103/PhysRevE.72.011408} {\bibfield  {journal} {\bibinfo  {journal} {Phys.
  Rev. E}\ }\textbf {\bibinfo {volume} {72}},\ \bibinfo {pages} {011408}
  (\bibinfo {year} {2005})}\BibitemShut {NoStop}%
\bibitem [{\citenamefont {Gompper}\ \emph {et~al.}(2009)\citenamefont
  {Gompper}, \citenamefont {Ihle}, \citenamefont {Kroll},\ and\ \citenamefont
  {Winkler}}]{gompper:2009}%
  \BibitemOpen
  \bibfield  {author} {\bibinfo {author} {\bibfnamefont {G.}~\bibnamefont
  {Gompper}}, \bibinfo {author} {\bibfnamefont {T.}~\bibnamefont {Ihle}},
  \bibinfo {author} {\bibfnamefont {D.}~\bibnamefont {Kroll}}, \ and\ \bibinfo
  {author} {\bibfnamefont {R.}~\bibnamefont {Winkler}},\ }in\ \href@noop {}
  {\emph {\bibinfo {booktitle} {Advanced computer simulation approaches for
  soft matter sciences III}}}\ (\bibinfo  {publisher} {Springer},\ \bibinfo
  {year} {2009})\ pp.\ \bibinfo {pages} {1--87}\BibitemShut {NoStop}%
\bibitem [{\citenamefont {Lobaskin}\ \emph {et~al.}(2004)\citenamefont
  {Lobaskin}, \citenamefont {D{\"u}nweg},\ and\ \citenamefont
  {Holm}}]{lobaskin:2004}%
  \BibitemOpen
  \bibfield  {author} {\bibinfo {author} {\bibfnamefont {V.}~\bibnamefont
  {Lobaskin}}, \bibinfo {author} {\bibfnamefont {B.}~\bibnamefont
  {D{\"u}nweg}}, \ and\ \bibinfo {author} {\bibfnamefont {C.}~\bibnamefont
  {Holm}},\ }\href@noop {} {\bibfield  {journal} {\bibinfo  {journal} {Journal
  of Physics: Condensed Matter}\ }\textbf {\bibinfo {volume} {16}},\ \bibinfo
  {pages} {S4063} (\bibinfo {year} {2004})}\BibitemShut {NoStop}%
\bibitem [{\citenamefont {Capuani}\ \emph {et~al.}(2004)\citenamefont
  {Capuani}, \citenamefont {Pagonabarraga},\ and\ \citenamefont
  {Frenkel}}]{capuani:2004}%
  \BibitemOpen
  \bibfield  {author} {\bibinfo {author} {\bibfnamefont {F.}~\bibnamefont
  {Capuani}}, \bibinfo {author} {\bibfnamefont {I.}~\bibnamefont
  {Pagonabarraga}}, \ and\ \bibinfo {author} {\bibfnamefont {D.}~\bibnamefont
  {Frenkel}},\ }\href {\doibase http://dx.doi.org/10.1063/1.1760739} {\bibfield
   {journal} {\bibinfo  {journal} {The Journal of Chemical Physics}\ }\textbf
  {\bibinfo {volume} {121}},\ \bibinfo {pages} {973} (\bibinfo {year}
  {2004})}\BibitemShut {NoStop}%
\bibitem [{\citenamefont {Benzi}\ \emph {et~al.}(1992)\citenamefont {Benzi},
  \citenamefont {Succi},\ and\ \citenamefont {Vergassola}}]{benzi:1992}%
  \BibitemOpen
  \bibfield  {author} {\bibinfo {author} {\bibfnamefont {R.}~\bibnamefont
  {Benzi}}, \bibinfo {author} {\bibfnamefont {S.}~\bibnamefont {Succi}}, \ and\
  \bibinfo {author} {\bibfnamefont {M.}~\bibnamefont {Vergassola}},\ }\href
  {\doibase http://dx.doi.org/10.1016/0370-1573(92)90090-M} {\bibfield
  {journal} {\bibinfo  {journal} {Physics Reports}\ }\textbf {\bibinfo {volume}
  {222}},\ \bibinfo {pages} {145 } (\bibinfo {year} {1992})}\BibitemShut
  {NoStop}%
\bibitem [{\citenamefont {Evans}(1979)}]{evans:1979}%
  \BibitemOpen
  \bibfield  {author} {\bibinfo {author} {\bibfnamefont {R.}~\bibnamefont
  {Evans}},\ }\href@noop {} {\bibfield  {journal} {\bibinfo  {journal}
  {Advances in Physics}\ }\textbf {\bibinfo {volume} {28}},\ \bibinfo {pages}
  {143} (\bibinfo {year} {1979})}\BibitemShut {NoStop}%
\bibitem [{\citenamefont {Rotenberg}\ \emph {et~al.}(2010)\citenamefont
  {Rotenberg}, \citenamefont {Pagonabarraga},\ and\ \citenamefont
  {Frenkel}}]{rotenberg:2010}%
  \BibitemOpen
  \bibfield  {author} {\bibinfo {author} {\bibfnamefont {B.}~\bibnamefont
  {Rotenberg}}, \bibinfo {author} {\bibfnamefont {I.}~\bibnamefont
  {Pagonabarraga}}, \ and\ \bibinfo {author} {\bibfnamefont {D.}~\bibnamefont
  {Frenkel}},\ }\href {\doibase 10.1039/B901553A} {\bibfield  {journal}
  {\bibinfo  {journal} {Faraday Discuss.}\ }\textbf {\bibinfo {volume} {144}},\
  \bibinfo {pages} {223} (\bibinfo {year} {2010})}\BibitemShut {NoStop}%
\bibitem [{\citenamefont {Pagonabarraga}\ \emph {et~al.}(2010)\citenamefont
  {Pagonabarraga}, \citenamefont {Rotenberg},\ and\ \citenamefont
  {Frenkel}}]{pagonabarraga:2010}%
  \BibitemOpen
  \bibfield  {author} {\bibinfo {author} {\bibfnamefont {I.}~\bibnamefont
  {Pagonabarraga}}, \bibinfo {author} {\bibfnamefont {B.}~\bibnamefont
  {Rotenberg}}, \ and\ \bibinfo {author} {\bibfnamefont {D.}~\bibnamefont
  {Frenkel}},\ }\href {\doibase 10.1039/C004012F} {\bibfield  {journal}
  {\bibinfo  {journal} {Phys. Chem. Chem. Phys.}\ }\textbf {\bibinfo {volume}
  {12}},\ \bibinfo {pages} {9566} (\bibinfo {year} {2010})}\BibitemShut
  {NoStop}%
\bibitem [{\citenamefont {Shan}\ and\ \citenamefont {Chen}(1993)}]{shan:1993}%
  \BibitemOpen
  \bibfield  {author} {\bibinfo {author} {\bibfnamefont {X.}~\bibnamefont
  {Shan}}\ and\ \bibinfo {author} {\bibfnamefont {H.}~\bibnamefont {Chen}},\
  }\href {\doibase 10.1103/PhysRevE.47.1815} {\bibfield  {journal} {\bibinfo
  {journal} {Phys. Rev. E}\ }\textbf {\bibinfo {volume} {47}},\ \bibinfo
  {pages} {1815} (\bibinfo {year} {1993})}\BibitemShut {NoStop}%
\bibitem [{\citenamefont {Shan}\ and\ \citenamefont {Chen}(1994)}]{shan:1994}%
  \BibitemOpen
  \bibfield  {author} {\bibinfo {author} {\bibfnamefont {X.}~\bibnamefont
  {Shan}}\ and\ \bibinfo {author} {\bibfnamefont {H.}~\bibnamefont {Chen}},\
  }\href {\doibase 10.1103/PhysRevE.49.2941} {\bibfield  {journal} {\bibinfo
  {journal} {Phys. Rev. E}\ }\textbf {\bibinfo {volume} {49}},\ \bibinfo
  {pages} {2941} (\bibinfo {year} {1994})}\BibitemShut {NoStop}%
\bibitem [{\citenamefont {Ladd}\ and\ \citenamefont
  {Verberg}(2001)}]{ladd:2001}%
  \BibitemOpen
  \bibfield  {author} {\bibinfo {author} {\bibfnamefont {A.~J.~C.}\
  \bibnamefont {Ladd}}\ and\ \bibinfo {author} {\bibfnamefont {R.}~\bibnamefont
  {Verberg}},\ }\href {\doibase 10.1023/A:1010414013942} {\bibfield  {journal}
  {\bibinfo  {journal} {Journal of Statistical Physics}\ }\textbf {\bibinfo
  {volume} {104}},\ \bibinfo {pages} {1191} (\bibinfo {year}
  {2001})}\BibitemShut {NoStop}%
\bibitem [{\citenamefont {Jansen}\ and\ \citenamefont
  {Harting}(2011)}]{jansen:2011}%
  \BibitemOpen
  \bibfield  {author} {\bibinfo {author} {\bibfnamefont {F.}~\bibnamefont
  {Jansen}}\ and\ \bibinfo {author} {\bibfnamefont {J.}~\bibnamefont
  {Harting}},\ }\href@noop {} {\bibfield  {journal} {\bibinfo  {journal}
  {Physical Review E}\ }\textbf {\bibinfo {volume} {83}},\ \bibinfo {pages}
  {046707} (\bibinfo {year} {2011})}\BibitemShut {NoStop}%
\bibitem [{\citenamefont {Kr{\"u}ger}\ \emph {et~al.}(2013)\citenamefont
  {Kr{\"u}ger}, \citenamefont {Frijters}, \citenamefont {G{\"u}nther},
  \citenamefont {Kaoui},\ and\ \citenamefont {Harting}}]{kruger:2013}%
  \BibitemOpen
  \bibfield  {author} {\bibinfo {author} {\bibfnamefont {T.}~\bibnamefont
  {Kr{\"u}ger}}, \bibinfo {author} {\bibfnamefont {S.}~\bibnamefont
  {Frijters}}, \bibinfo {author} {\bibfnamefont {F.}~\bibnamefont
  {G{\"u}nther}}, \bibinfo {author} {\bibfnamefont {B.}~\bibnamefont {Kaoui}},
  \ and\ \bibinfo {author} {\bibfnamefont {J.}~\bibnamefont {Harting}},\
  }\href@noop {} {\bibfield  {journal} {\bibinfo  {journal} {The European
  Physical Journal Special Topics}\ }\textbf {\bibinfo {volume} {222}},\
  \bibinfo {pages} {177} (\bibinfo {year} {2013})}\BibitemShut {NoStop}%
\bibitem [{\citenamefont {Kuron}\ \emph {et~al.}(2016)\citenamefont {Kuron},
  \citenamefont {Rempfer}, \citenamefont {Schornbaum}, \citenamefont {Bauer},
  \citenamefont {Godenschwager}, \citenamefont {Holm},\ and\ \citenamefont
  {de~Graaf}}]{kuron:2016}%
  \BibitemOpen
  \bibfield  {author} {\bibinfo {author} {\bibfnamefont {M.}~\bibnamefont
  {Kuron}}, \bibinfo {author} {\bibfnamefont {G.}~\bibnamefont {Rempfer}},
  \bibinfo {author} {\bibfnamefont {F.}~\bibnamefont {Schornbaum}}, \bibinfo
  {author} {\bibfnamefont {M.}~\bibnamefont {Bauer}}, \bibinfo {author}
  {\bibfnamefont {C.}~\bibnamefont {Godenschwager}}, \bibinfo {author}
  {\bibfnamefont {C.}~\bibnamefont {Holm}}, \ and\ \bibinfo {author}
  {\bibfnamefont {J.}~\bibnamefont {de~Graaf}},\ }\href@noop {} {\bibfield
  {journal} {\bibinfo  {journal} {The Journal of Chemical Physics}\ }\textbf
  {\bibinfo {volume} {145}},\ \bibinfo {pages} {214102} (\bibinfo {year}
  {2016})}\BibitemShut {NoStop}%
\bibitem [{\citenamefont {Bhatnagar}\ \emph {et~al.}(1954)\citenamefont
  {Bhatnagar}, \citenamefont {Gross},\ and\ \citenamefont
  {Krook}}]{bhatnagar:1954}%
  \BibitemOpen
  \bibfield  {author} {\bibinfo {author} {\bibfnamefont {P.~L.}\ \bibnamefont
  {Bhatnagar}}, \bibinfo {author} {\bibfnamefont {E.~P.}\ \bibnamefont
  {Gross}}, \ and\ \bibinfo {author} {\bibfnamefont {M.}~\bibnamefont
  {Krook}},\ }\href {\doibase 10.1103/PhysRev.94.511} {\bibfield  {journal}
  {\bibinfo  {journal} {Phys. Rev.}\ }\textbf {\bibinfo {volume} {94}},\
  \bibinfo {pages} {511} (\bibinfo {year} {1954})}\BibitemShut {NoStop}%
\bibitem [{\citenamefont {Qian}\ \emph {et~al.}(1992)\citenamefont {Qian},
  \citenamefont {D'Humieres},\ and\ \citenamefont {Lallemand}}]{qian:1992}%
  \BibitemOpen
  \bibfield  {author} {\bibinfo {author} {\bibfnamefont {Y.~H.}\ \bibnamefont
  {Qian}}, \bibinfo {author} {\bibfnamefont {D.}~\bibnamefont {D'Humieres}}, \
  and\ \bibinfo {author} {\bibfnamefont {P.}~\bibnamefont {Lallemand}},\ }\href
  {http://stacks.iop.org/0295-5075/17/i=6/a=001} {\bibfield  {journal}
  {\bibinfo  {journal} {EPL (Europhysics Letters)}\ }\textbf {\bibinfo {volume}
  {17}},\ \bibinfo {pages} {479} (\bibinfo {year} {1992})}\BibitemShut
  {NoStop}%
\bibitem [{\citenamefont {Greberg}\ and\ \citenamefont
  {Kjellander}(1998)}]{greberg1998charge}%
  \BibitemOpen
  \bibfield  {author} {\bibinfo {author} {\bibfnamefont {H.}~\bibnamefont
  {Greberg}}\ and\ \bibinfo {author} {\bibfnamefont {R.}~\bibnamefont
  {Kjellander}},\ }\href@noop {} {\bibfield  {journal} {\bibinfo  {journal}
  {The Journal of chemical physics}\ }\textbf {\bibinfo {volume} {108}},\
  \bibinfo {pages} {2940} (\bibinfo {year} {1998})}\BibitemShut {NoStop}%
\bibitem [{\citenamefont {Hunter}(2003)}]{hunter2003significance}%
  \BibitemOpen
  \bibfield  {author} {\bibinfo {author} {\bibfnamefont {R.~J.}\ \bibnamefont
  {Hunter}},\ }\href@noop {} {\bibfield  {journal} {\bibinfo  {journal}
  {Advances in colloid and interface science}\ }\textbf {\bibinfo {volume}
  {100}},\ \bibinfo {pages} {153} (\bibinfo {year} {2003})}\BibitemShut
  {NoStop}%
\bibitem [{\citenamefont {Qiao}\ and\ \citenamefont
  {Aluru}(2004)}]{qiao2004charge}%
  \BibitemOpen
  \bibfield  {author} {\bibinfo {author} {\bibfnamefont {R.}~\bibnamefont
  {Qiao}}\ and\ \bibinfo {author} {\bibfnamefont {N.}~\bibnamefont {Aluru}},\
  }\href@noop {} {\bibfield  {journal} {\bibinfo  {journal} {Physical review
  letters}\ }\textbf {\bibinfo {volume} {92}},\ \bibinfo {pages} {198301}
  (\bibinfo {year} {2004})}\BibitemShut {NoStop}%
\bibitem [{\citenamefont {Van~der Heyden}\ \emph {et~al.}(2006)\citenamefont
  {Van~der Heyden}, \citenamefont {Stein}, \citenamefont {Besteman},
  \citenamefont {Lemay},\ and\ \citenamefont {Dekker}}]{van2006charge}%
  \BibitemOpen
  \bibfield  {author} {\bibinfo {author} {\bibfnamefont {F.~H.}\ \bibnamefont
  {Van~der Heyden}}, \bibinfo {author} {\bibfnamefont {D.}~\bibnamefont
  {Stein}}, \bibinfo {author} {\bibfnamefont {K.}~\bibnamefont {Besteman}},
  \bibinfo {author} {\bibfnamefont {S.~G.}\ \bibnamefont {Lemay}}, \ and\
  \bibinfo {author} {\bibfnamefont {C.}~\bibnamefont {Dekker}},\ }\href@noop {}
  {\bibfield  {journal} {\bibinfo  {journal} {Physical review letters}\
  }\textbf {\bibinfo {volume} {96}},\ \bibinfo {pages} {224502} (\bibinfo
  {year} {2006})}\BibitemShut {NoStop}%
\bibitem [{\citenamefont {Ramadugu}\ \emph {et~al.}(2013)\citenamefont
  {Ramadugu}, \citenamefont {Thampi}, \citenamefont {Adhikari}, \citenamefont
  {Succi},\ and\ \citenamefont {Ansumali}}]{ramadugu2013lattice}%
  \BibitemOpen
  \bibfield  {author} {\bibinfo {author} {\bibfnamefont {R.}~\bibnamefont
  {Ramadugu}}, \bibinfo {author} {\bibfnamefont {S.~P.}\ \bibnamefont
  {Thampi}}, \bibinfo {author} {\bibfnamefont {R.}~\bibnamefont {Adhikari}},
  \bibinfo {author} {\bibfnamefont {S.}~\bibnamefont {Succi}}, \ and\ \bibinfo
  {author} {\bibfnamefont {S.}~\bibnamefont {Ansumali}},\ }\href@noop {}
  {\bibfield  {journal} {\bibinfo  {journal} {EPL (Europhysics Letters)}\
  }\textbf {\bibinfo {volume} {101}},\ \bibinfo {pages} {50006} (\bibinfo
  {year} {2013})}\BibitemShut {NoStop}%
\bibitem [{\citenamefont {Rempfer}\ \emph {et~al.}(2016)\citenamefont
  {Rempfer}, \citenamefont {Davies}, \citenamefont {Holm},\ and\ \citenamefont
  {de~Graaf}}]{rempfer:2016}%
  \BibitemOpen
  \bibfield  {author} {\bibinfo {author} {\bibfnamefont {G.}~\bibnamefont
  {Rempfer}}, \bibinfo {author} {\bibfnamefont {G.~B.}\ \bibnamefont {Davies}},
  \bibinfo {author} {\bibfnamefont {C.}~\bibnamefont {Holm}}, \ and\ \bibinfo
  {author} {\bibfnamefont {J.}~\bibnamefont {de~Graaf}},\ }\href@noop {}
  {\bibfield  {journal} {\bibinfo  {journal} {The Journal of Chemical Physics}\
  }\textbf {\bibinfo {volume} {145}},\ \bibinfo {pages} {044901} (\bibinfo
  {year} {2016})}\BibitemShut {NoStop}%
\bibitem [{\citenamefont {Liu}\ \emph {et~al.}(2016)\citenamefont {Liu},
  \citenamefont {Kang}, \citenamefont {Leonardi}, \citenamefont {Schmieschek},
  \citenamefont {Narv{\'a}ez}, \citenamefont {Jones}, \citenamefont {Williams},
  \citenamefont {Valocchi},\ and\ \citenamefont {Harting}}]{liu:2016}%
  \BibitemOpen
  \bibfield  {author} {\bibinfo {author} {\bibfnamefont {H.}~\bibnamefont
  {Liu}}, \bibinfo {author} {\bibfnamefont {Q.}~\bibnamefont {Kang}}, \bibinfo
  {author} {\bibfnamefont {C.~R.}\ \bibnamefont {Leonardi}}, \bibinfo {author}
  {\bibfnamefont {S.}~\bibnamefont {Schmieschek}}, \bibinfo {author}
  {\bibfnamefont {A.}~\bibnamefont {Narv{\'a}ez}}, \bibinfo {author}
  {\bibfnamefont {B.~D.}\ \bibnamefont {Jones}}, \bibinfo {author}
  {\bibfnamefont {J.~R.}\ \bibnamefont {Williams}}, \bibinfo {author}
  {\bibfnamefont {A.~J.}\ \bibnamefont {Valocchi}}, \ and\ \bibinfo {author}
  {\bibfnamefont {J.}~\bibnamefont {Harting}},\ }\href {\doibase
  10.1007/s10596-015-9542-3} {\bibfield  {journal} {\bibinfo  {journal}
  {Computational Geosciences}\ }\textbf {\bibinfo {volume} {20}},\ \bibinfo
  {pages} {777} (\bibinfo {year} {2016})}\BibitemShut {NoStop}%
\bibitem [{\citenamefont {Sbragaglia}\ \emph {et~al.}(2009)\citenamefont
  {Sbragaglia}, \citenamefont {Chen}, \citenamefont {Shan},\ and\ \citenamefont
  {Succi}}]{sbragaglia2009continuum}%
  \BibitemOpen
  \bibfield  {author} {\bibinfo {author} {\bibfnamefont {M.}~\bibnamefont
  {Sbragaglia}}, \bibinfo {author} {\bibfnamefont {H.}~\bibnamefont {Chen}},
  \bibinfo {author} {\bibfnamefont {X.}~\bibnamefont {Shan}}, \ and\ \bibinfo
  {author} {\bibfnamefont {S.}~\bibnamefont {Succi}},\ }\href@noop {}
  {\bibfield  {journal} {\bibinfo  {journal} {EPL (Europhysics Letters)}\
  }\textbf {\bibinfo {volume} {86}},\ \bibinfo {pages} {24005} (\bibinfo {year}
  {2009})}\BibitemShut {NoStop}%
\bibitem [{\citenamefont {Onuki}(2006{\natexlab{a}})}]{onuki:2006}%
  \BibitemOpen
  \bibfield  {author} {\bibinfo {author} {\bibfnamefont {A.}~\bibnamefont
  {Onuki}},\ }\href {\doibase 10.1103/PhysRevE.73.021506} {\bibfield  {journal}
  {\bibinfo  {journal} {Phys. Rev. E}\ }\textbf {\bibinfo {volume} {73}},\
  \bibinfo {pages} {021506} (\bibinfo {year} {2006}{\natexlab{a}})}\BibitemShut
  {NoStop}%
\bibitem [{\citenamefont {Onuki}\ \emph {et~al.}(2011)\citenamefont {Onuki},
  \citenamefont {Okamoto},\ and\ \citenamefont {Araki}}]{onuki2011phase}%
  \BibitemOpen
  \bibfield  {author} {\bibinfo {author} {\bibfnamefont {A.}~\bibnamefont
  {Onuki}}, \bibinfo {author} {\bibfnamefont {R.}~\bibnamefont {Okamoto}}, \
  and\ \bibinfo {author} {\bibfnamefont {T.}~\bibnamefont {Araki}},\
  }\href@noop {} {\bibfield  {journal} {\bibinfo  {journal} {Bulletin of the
  Chemical Society of Japan}\ }\textbf {\bibinfo {volume} {84}},\ \bibinfo
  {pages} {569} (\bibinfo {year} {2011})}\BibitemShut {NoStop}%
\bibitem [{\citenamefont {Landau}\ and\ \citenamefont
  {Lifshitz}(1960)}]{landau1960electrodynamics}%
  \BibitemOpen
  \bibfield  {author} {\bibinfo {author} {\bibfnamefont {L.~D.}\ \bibnamefont
  {Landau}}\ and\ \bibinfo {author} {\bibfnamefont {E.~M.}\ \bibnamefont
  {Lifshitz}},\ }\href@noop {} {\emph {\bibinfo {title} {Electrodynamics of
  Continuous Media}}}\ (\bibinfo  {publisher} {Pergamon Press, Oxford},\
  \bibinfo {year} {1960})\BibitemShut {NoStop}%
\bibitem [{\citenamefont {Zahn}(2006)}]{zahn:2006}%
  \BibitemOpen
  \bibfield  {author} {\bibinfo {author} {\bibfnamefont {M.}~\bibnamefont
  {Zahn}},\ }in\ \href {\doibase 10.1109/CEIDP.2006.312092} {\emph {\bibinfo
  {booktitle} {2006 IEEE Conference on Electrical Insulation and Dielectric
  Phenomena}}}\ (\bibinfo {year} {2006})\ pp.\ \bibinfo {pages}
  {186--189}\BibitemShut {NoStop}%
\bibitem [{\citenamefont {Oettel}\ \emph {et~al.}(2005)\citenamefont {Oettel},
  \citenamefont {Dominguez},\ and\ \citenamefont
  {Dietrich}}]{oettel2005attractions}%
  \BibitemOpen
  \bibfield  {author} {\bibinfo {author} {\bibfnamefont {M.}~\bibnamefont
  {Oettel}}, \bibinfo {author} {\bibfnamefont {A.}~\bibnamefont {Dominguez}}, \
  and\ \bibinfo {author} {\bibfnamefont {S.}~\bibnamefont {Dietrich}},\
  }\href@noop {} {\bibfield  {journal} {\bibinfo  {journal} {Journal of
  Physics: Condensed Matter}\ }\textbf {\bibinfo {volume} {17}},\ \bibinfo
  {pages} {L337} (\bibinfo {year} {2005})}\BibitemShut {NoStop}%
\bibitem [{\citenamefont {Dominguez}\ \emph {et~al.}(2008)\citenamefont
  {Dominguez}, \citenamefont {Frydel},\ and\ \citenamefont
  {Oettel}}]{dominguez2008multipole}%
  \BibitemOpen
  \bibfield  {author} {\bibinfo {author} {\bibfnamefont {A.}~\bibnamefont
  {Dominguez}}, \bibinfo {author} {\bibfnamefont {D.}~\bibnamefont {Frydel}}, \
  and\ \bibinfo {author} {\bibfnamefont {M.}~\bibnamefont {Oettel}},\
  }\href@noop {} {\bibfield  {journal} {\bibinfo  {journal} {Physical Review
  E}\ }\textbf {\bibinfo {volume} {77}},\ \bibinfo {pages} {020401} (\bibinfo
  {year} {2008})}\BibitemShut {NoStop}%
\bibitem [{\citenamefont {Majee}\ \emph {et~al.}(2014)\citenamefont {Majee},
  \citenamefont {Bier},\ and\ \citenamefont
  {Dietrich}}]{majee2014electrostatic}%
  \BibitemOpen
  \bibfield  {author} {\bibinfo {author} {\bibfnamefont {A.}~\bibnamefont
  {Majee}}, \bibinfo {author} {\bibfnamefont {M.}~\bibnamefont {Bier}}, \ and\
  \bibinfo {author} {\bibfnamefont {S.}~\bibnamefont {Dietrich}},\ }\href@noop
  {} {\bibfield  {journal} {\bibinfo  {journal} {The Journal of Chemical
  Physics}\ }\textbf {\bibinfo {volume} {140}},\ \bibinfo {pages} {164906}
  (\bibinfo {year} {2014})}\BibitemShut {NoStop}%
\bibitem [{\citenamefont {Girotto}\ \emph {et~al.}(2015)\citenamefont
  {Girotto}, \citenamefont {dos Santos},\ and\ \citenamefont
  {Levin}}]{girotto2015interaction}%
  \BibitemOpen
  \bibfield  {author} {\bibinfo {author} {\bibfnamefont {M.}~\bibnamefont
  {Girotto}}, \bibinfo {author} {\bibfnamefont {A.~P.}\ \bibnamefont {dos
  Santos}}, \ and\ \bibinfo {author} {\bibfnamefont {Y.}~\bibnamefont
  {Levin}},\ }\href@noop {} {\bibfield  {journal} {\bibinfo  {journal} {The
  Journal of Physical Chemistry B}\ }\textbf {\bibinfo {volume} {120}},\
  \bibinfo {pages} {5817} (\bibinfo {year} {2015})}\BibitemShut {NoStop}%
\bibitem [{\citenamefont {Majee}\ \emph {et~al.}(2016)\citenamefont {Majee},
  \citenamefont {Bier},\ and\ \citenamefont {Dietrich}}]{majee2016poisson}%
  \BibitemOpen
  \bibfield  {author} {\bibinfo {author} {\bibfnamefont {A.}~\bibnamefont
  {Majee}}, \bibinfo {author} {\bibfnamefont {M.}~\bibnamefont {Bier}}, \ and\
  \bibinfo {author} {\bibfnamefont {S.}~\bibnamefont {Dietrich}},\ }\href@noop
  {} {\bibfield  {journal} {\bibinfo  {journal} {The Journal of Chemical
  Physics}\ }\textbf {\bibinfo {volume} {145}},\ \bibinfo {pages} {064707}
  (\bibinfo {year} {2016})}\BibitemShut {NoStop}%
\bibitem [{\citenamefont {Aidun}\ \emph {et~al.}(1998)\citenamefont {Aidun},
  \citenamefont {Lu},\ and\ \citenamefont {Ding}}]{aidun:1998}%
  \BibitemOpen
  \bibfield  {author} {\bibinfo {author} {\bibfnamefont {C.~K.}\ \bibnamefont
  {Aidun}}, \bibinfo {author} {\bibfnamefont {Y.}~\bibnamefont {Lu}}, \ and\
  \bibinfo {author} {\bibfnamefont {E.-J.}\ \bibnamefont {Ding}},\ }\href@noop
  {} {\bibfield  {journal} {\bibinfo  {journal} {Journal of Fluid Mechanics}\
  }\textbf {\bibinfo {volume} {373}},\ \bibinfo {pages} {287} (\bibinfo {year}
  {1998})}\BibitemShut {NoStop}%
\bibitem [{\citenamefont {Noble}\ and\ \citenamefont
  {Torczynski}(1998)}]{noble:1998}%
  \BibitemOpen
  \bibfield  {author} {\bibinfo {author} {\bibfnamefont {D.~R.}\ \bibnamefont
  {Noble}}\ and\ \bibinfo {author} {\bibfnamefont {J.~R.}\ \bibnamefont
  {Torczynski}},\ }\href {\doibase 10.1142/S0129183198001084} {\bibfield
  {journal} {\bibinfo  {journal} {International Journal of Modern Physics C}\
  }\textbf {\bibinfo {volume} {09}},\ \bibinfo {pages} {1189} (\bibinfo {year}
  {1998})}\BibitemShut {NoStop}%
\bibitem [{\citenamefont {Onuki}(2006{\natexlab{b}})}]{onuki2006ginzburg}%
  \BibitemOpen
  \bibfield  {author} {\bibinfo {author} {\bibfnamefont {A.}~\bibnamefont
  {Onuki}},\ }\href@noop {} {\bibfield  {journal} {\bibinfo  {journal}
  {Physical Review E}\ }\textbf {\bibinfo {volume} {73}},\ \bibinfo {pages}
  {021506} (\bibinfo {year} {2006}{\natexlab{b}})}\BibitemShut {NoStop}%
\bibitem [{\citenamefont {Bier}\ \emph {et~al.}(2008)\citenamefont {Bier},
  \citenamefont {Zwanikken},\ and\ \citenamefont {van Roij}}]{bier:2008}%
  \BibitemOpen
  \bibfield  {author} {\bibinfo {author} {\bibfnamefont {M.}~\bibnamefont
  {Bier}}, \bibinfo {author} {\bibfnamefont {J.}~\bibnamefont {Zwanikken}}, \
  and\ \bibinfo {author} {\bibfnamefont {R.}~\bibnamefont {van Roij}},\ }\href
  {\doibase 10.1103/PhysRevLett.101.046104} {\bibfield  {journal} {\bibinfo
  {journal} {Phys. Rev. Lett.}\ }\textbf {\bibinfo {volume} {101}},\ \bibinfo
  {pages} {046104} (\bibinfo {year} {2008})}\BibitemShut {NoStop}%
\bibitem [{\citenamefont {O'Konski}\ and\ \citenamefont
  {Thacher~Jr}(1953)}]{okonski:1953}%
  \BibitemOpen
  \bibfield  {author} {\bibinfo {author} {\bibfnamefont {C.~T.}\ \bibnamefont
  {O'Konski}}\ and\ \bibinfo {author} {\bibfnamefont {H.~C.}\ \bibnamefont
  {Thacher~Jr}},\ }\href@noop {} {\bibfield  {journal} {\bibinfo  {journal}
  {The Journal of Physical Chemistry}\ }\textbf {\bibinfo {volume} {57}},\
  \bibinfo {pages} {955} (\bibinfo {year} {1953})}\BibitemShut {NoStop}%
\bibitem [{\citenamefont {Stone}(1994)}]{stone:1994}%
  \BibitemOpen
  \bibfield  {author} {\bibinfo {author} {\bibfnamefont {H.~A.}\ \bibnamefont
  {Stone}},\ }\href@noop {} {\bibfield  {journal} {\bibinfo  {journal} {Annual
  Review of Fluid Mechanics}\ }\textbf {\bibinfo {volume} {26}},\ \bibinfo
  {pages} {65} (\bibinfo {year} {1994})}\BibitemShut {NoStop}%
\bibitem [{\citenamefont {Nganguia}\ \emph {et~al.}(2016)\citenamefont
  {Nganguia}, \citenamefont {Young}, \citenamefont {Layton}, \citenamefont
  {Lai},\ and\ \citenamefont {Hu}}]{nganguia:2016}%
  \BibitemOpen
  \bibfield  {author} {\bibinfo {author} {\bibfnamefont {H.}~\bibnamefont
  {Nganguia}}, \bibinfo {author} {\bibfnamefont {Y.-N.}\ \bibnamefont {Young}},
  \bibinfo {author} {\bibfnamefont {A.}~\bibnamefont {Layton}}, \bibinfo
  {author} {\bibfnamefont {M.-C.}\ \bibnamefont {Lai}}, \ and\ \bibinfo
  {author} {\bibfnamefont {W.-F.}\ \bibnamefont {Hu}},\ }\href@noop {}
  {\bibfield  {journal} {\bibinfo  {journal} {Physical Review E}\ }\textbf
  {\bibinfo {volume} {93}},\ \bibinfo {pages} {053114} (\bibinfo {year}
  {2016})}\BibitemShut {NoStop}%
\bibitem [{\citenamefont {Ha}\ and\ \citenamefont {Yang}(1998)}]{ha:1998}%
  \BibitemOpen
  \bibfield  {author} {\bibinfo {author} {\bibfnamefont {J.-W.}\ \bibnamefont
  {Ha}}\ and\ \bibinfo {author} {\bibfnamefont {S.-M.}\ \bibnamefont {Yang}},\
  }\href@noop {} {\bibfield  {journal} {\bibinfo  {journal} {Journal of colloid
  and interface science}\ }\textbf {\bibinfo {volume} {206}},\ \bibinfo {pages}
  {195} (\bibinfo {year} {1998})}\BibitemShut {NoStop}%
\bibitem [{\citenamefont {Lac}\ and\ \citenamefont {Homsy}(2007)}]{lac:2007}%
  \BibitemOpen
  \bibfield  {author} {\bibinfo {author} {\bibfnamefont {E.}~\bibnamefont
  {Lac}}\ and\ \bibinfo {author} {\bibfnamefont {G.}~\bibnamefont {Homsy}},\
  }\href@noop {} {\bibfield  {journal} {\bibinfo  {journal} {Journal of Fluid
  Mechanics}\ }\textbf {\bibinfo {volume} {590}},\ \bibinfo {pages} {239}
  (\bibinfo {year} {2007})}\BibitemShut {NoStop}%
\bibitem [{\citenamefont {Supeene}\ \emph {et~al.}(2008)\citenamefont
  {Supeene}, \citenamefont {Koch},\ and\ \citenamefont
  {Bhattacharjee}}]{supeene:2008}%
  \BibitemOpen
  \bibfield  {author} {\bibinfo {author} {\bibfnamefont {G.}~\bibnamefont
  {Supeene}}, \bibinfo {author} {\bibfnamefont {C.~R.}\ \bibnamefont {Koch}}, \
  and\ \bibinfo {author} {\bibfnamefont {S.}~\bibnamefont {Bhattacharjee}},\
  }\href@noop {} {\bibfield  {journal} {\bibinfo  {journal} {Journal of colloid
  and interface science}\ }\textbf {\bibinfo {volume} {318}},\ \bibinfo {pages}
  {463} (\bibinfo {year} {2008})}\BibitemShut {NoStop}%
\bibitem [{\citenamefont {Nganguia}\ \emph {et~al.}(2015)\citenamefont
  {Nganguia}, \citenamefont {Young}, \citenamefont {Layton}, \citenamefont
  {Hu},\ and\ \citenamefont {Lai}}]{nganguia:2015}%
  \BibitemOpen
  \bibfield  {author} {\bibinfo {author} {\bibfnamefont {H.}~\bibnamefont
  {Nganguia}}, \bibinfo {author} {\bibfnamefont {Y.-N.}\ \bibnamefont {Young}},
  \bibinfo {author} {\bibfnamefont {A.}~\bibnamefont {Layton}}, \bibinfo
  {author} {\bibfnamefont {W.-F.}\ \bibnamefont {Hu}}, \ and\ \bibinfo {author}
  {\bibfnamefont {M.-C.}\ \bibnamefont {Lai}},\ }\href@noop {} {\bibfield
  {journal} {\bibinfo  {journal} {Communications in Computational Physics}\
  }\textbf {\bibinfo {volume} {18}},\ \bibinfo {pages} {429} (\bibinfo {year}
  {2015})}\BibitemShut {NoStop}%
\bibitem [{\citenamefont {Pillai}\ \emph {et~al.}(2016)\citenamefont {Pillai},
  \citenamefont {Berry}, \citenamefont {Harvie},\ and\ \citenamefont
  {Davidson}}]{pillai:2016}%
  \BibitemOpen
  \bibfield  {author} {\bibinfo {author} {\bibfnamefont {R.}~\bibnamefont
  {Pillai}}, \bibinfo {author} {\bibfnamefont {J.~D.}\ \bibnamefont {Berry}},
  \bibinfo {author} {\bibfnamefont {D.~J.}\ \bibnamefont {Harvie}}, \ and\
  \bibinfo {author} {\bibfnamefont {M.~R.}\ \bibnamefont {Davidson}},\
  }\href@noop {} {\bibfield  {journal} {\bibinfo  {journal} {Soft matter}\
  }\textbf {\bibinfo {volume} {12}},\ \bibinfo {pages} {3310} (\bibinfo {year}
  {2016})}\BibitemShut {NoStop}%
\bibitem [{\citenamefont {Komrakova}\ \emph {et~al.}(2014)\citenamefont
  {Komrakova}, \citenamefont {Shardt}, \citenamefont {Eskin},\ and\
  \citenamefont {Derksen}}]{komrakova2014lattice}%
  \BibitemOpen
  \bibfield  {author} {\bibinfo {author} {\bibfnamefont {A.}~\bibnamefont
  {Komrakova}}, \bibinfo {author} {\bibfnamefont {O.}~\bibnamefont {Shardt}},
  \bibinfo {author} {\bibfnamefont {D.}~\bibnamefont {Eskin}}, \ and\ \bibinfo
  {author} {\bibfnamefont {J.}~\bibnamefont {Derksen}},\ }\href@noop {}
  {\bibfield  {journal} {\bibinfo  {journal} {International Journal of
  Multiphase Flow}\ }\textbf {\bibinfo {volume} {59}},\ \bibinfo {pages} {24}
  (\bibinfo {year} {2014})}\BibitemShut {NoStop}%
\bibitem [{\citenamefont {Kulkarni}\ and\ \citenamefont
  {Sojka}(2014)}]{kulkarni:2014}%
  \BibitemOpen
  \bibfield  {author} {\bibinfo {author} {\bibfnamefont {V.}~\bibnamefont
  {Kulkarni}}\ and\ \bibinfo {author} {\bibfnamefont {P.}~\bibnamefont
  {Sojka}},\ }\href@noop {} {\bibfield  {journal} {\bibinfo  {journal} {Physics
  of Fluids}\ }\textbf {\bibinfo {volume} {26}},\ \bibinfo {pages} {072103}
  (\bibinfo {year} {2014})}\BibitemShut {NoStop}%
\bibitem [{\citenamefont {Guildenbecher}\ \emph {et~al.}(2009)\citenamefont
  {Guildenbecher}, \citenamefont {L{\'o}pez-Rivera},\ and\ \citenamefont
  {Sojka}}]{guildenbecher:2009}%
  \BibitemOpen
  \bibfield  {author} {\bibinfo {author} {\bibfnamefont {D.~R.}\ \bibnamefont
  {Guildenbecher}}, \bibinfo {author} {\bibfnamefont {C.}~\bibnamefont
  {L{\'o}pez-Rivera}}, \ and\ \bibinfo {author} {\bibfnamefont {P.~E.}\
  \bibnamefont {Sojka}},\ }\href {\doibase 10.1007/s00348-008-0593-2}
  {\bibfield  {journal} {\bibinfo  {journal} {Experiments in Fluids}\ }\textbf
  {\bibinfo {volume} {46}},\ \bibinfo {pages} {371} (\bibinfo {year}
  {2009})}\BibitemShut {NoStop}%
\bibitem [{\citenamefont {Gelfand}(1996)}]{gelfand:1996}%
  \BibitemOpen
  \bibfield  {author} {\bibinfo {author} {\bibfnamefont {B.}~\bibnamefont
  {Gelfand}},\ }\href@noop {} {\bibfield  {journal} {\bibinfo  {journal}
  {Progress in energy and combustion science}\ }\textbf {\bibinfo {volume}
  {22}},\ \bibinfo {pages} {201} (\bibinfo {year} {1996})}\BibitemShut
  {NoStop}%
\bibitem [{\citenamefont {Lobaskin}\ \emph {et~al.}(2007)\citenamefont
  {Lobaskin}, \citenamefont {D\"unweg}, \citenamefont {Medebach}, \citenamefont
  {Palberg},\ and\ \citenamefont {Holm}}]{lobaskin:2007}%
  \BibitemOpen
  \bibfield  {author} {\bibinfo {author} {\bibfnamefont {V.}~\bibnamefont
  {Lobaskin}}, \bibinfo {author} {\bibfnamefont {B.}~\bibnamefont {D\"unweg}},
  \bibinfo {author} {\bibfnamefont {M.}~\bibnamefont {Medebach}}, \bibinfo
  {author} {\bibfnamefont {T.}~\bibnamefont {Palberg}}, \ and\ \bibinfo
  {author} {\bibfnamefont {C.}~\bibnamefont {Holm}},\ }\href {\doibase
  10.1103/PhysRevLett.98.176105} {\bibfield  {journal} {\bibinfo  {journal}
  {Phys. Rev. Lett.}\ }\textbf {\bibinfo {volume} {98}},\ \bibinfo {pages}
  {176105} (\bibinfo {year} {2007})}\BibitemShut {NoStop}%
\bibitem [{\citenamefont {Giupponi}\ and\ \citenamefont
  {Pagonabarraga}(2011)}]{giupponi:2011}%
  \BibitemOpen
  \bibfield  {author} {\bibinfo {author} {\bibfnamefont {G.}~\bibnamefont
  {Giupponi}}\ and\ \bibinfo {author} {\bibfnamefont {I.}~\bibnamefont
  {Pagonabarraga}},\ }\href@noop {} {\bibfield  {journal} {\bibinfo  {journal}
  {Physical review letters}\ }\textbf {\bibinfo {volume} {106}},\ \bibinfo
  {pages} {248304} (\bibinfo {year} {2011})}\BibitemShut {NoStop}%
\bibitem [{\citenamefont {O{'}Brien}\ and\ \citenamefont
  {White}(1978)}]{obrien:1978}%
  \BibitemOpen
  \bibfield  {author} {\bibinfo {author} {\bibfnamefont {R.~W.}\ \bibnamefont
  {O{'}Brien}}\ and\ \bibinfo {author} {\bibfnamefont {L.~R.}\ \bibnamefont
  {White}},\ }\href {\doibase 10.1039/F29787401607} {\bibfield  {journal}
  {\bibinfo  {journal} {J. Chem. Soc.{,} Faraday Trans. 2}\ }\textbf {\bibinfo
  {volume} {74}},\ \bibinfo {pages} {1607} (\bibinfo {year}
  {1978})}\BibitemShut {NoStop}%
\bibitem [{\citenamefont {Caruso}\ and\ \citenamefont
  {Antonietti}(2001)}]{caruso:2001}%
  \BibitemOpen
  \bibfield  {author} {\bibinfo {author} {\bibfnamefont {R.~A.}\ \bibnamefont
  {Caruso}}\ and\ \bibinfo {author} {\bibfnamefont {M.}~\bibnamefont
  {Antonietti}},\ }\href@noop {} {\bibfield  {journal} {\bibinfo  {journal}
  {Chemistry of materials}\ }\textbf {\bibinfo {volume} {13}},\ \bibinfo
  {pages} {3272} (\bibinfo {year} {2001})}\BibitemShut {NoStop}%
\bibitem [{\citenamefont {Gurrappa}\ and\ \citenamefont
  {Binder}(2008)}]{gurrappa:2008}%
  \BibitemOpen
  \bibfield  {author} {\bibinfo {author} {\bibfnamefont {I.}~\bibnamefont
  {Gurrappa}}\ and\ \bibinfo {author} {\bibfnamefont {L.}~\bibnamefont
  {Binder}},\ }\href@noop {} {\bibfield  {journal} {\bibinfo  {journal}
  {Science and Technology of Advanced Materials}\ }\textbf {\bibinfo {volume}
  {9}},\ \bibinfo {pages} {043001} (\bibinfo {year} {2008})}\BibitemShut
  {NoStop}%
\bibitem [{\citenamefont {Chevalier}\ and\ \citenamefont
  {Bolzinger}(2013)}]{chevalier:2013}%
  \BibitemOpen
  \bibfield  {author} {\bibinfo {author} {\bibfnamefont {Y.}~\bibnamefont
  {Chevalier}}\ and\ \bibinfo {author} {\bibfnamefont {M.-A.}\ \bibnamefont
  {Bolzinger}},\ }\href@noop {} {\bibfield  {journal} {\bibinfo  {journal}
  {Colloids and Surfaces A: Physicochemical and Engineering Aspects}\ }\textbf
  {\bibinfo {volume} {439}},\ \bibinfo {pages} {23} (\bibinfo {year}
  {2013})}\BibitemShut {NoStop}%
\bibitem [{\citenamefont {Chang}\ and\ \citenamefont {Yeo}(2010)}]{chang:2010}%
  \BibitemOpen
  \bibfield  {author} {\bibinfo {author} {\bibfnamefont {H.-C.}\ \bibnamefont
  {Chang}}\ and\ \bibinfo {author} {\bibfnamefont {L.~Y.}\ \bibnamefont
  {Yeo}},\ }\href@noop {} {\emph {\bibinfo {title} {Electrokinetically driven
  microfluidics and nanofluidics}}}\ (\bibinfo  {publisher} {Cambridge
  University Press New York},\ \bibinfo {year} {2010})\BibitemShut {NoStop}%
\bibitem [{\citenamefont {Pieranski}(1980)}]{pieranski1980two}%
  \BibitemOpen
  \bibfield  {author} {\bibinfo {author} {\bibfnamefont {P.}~\bibnamefont
  {Pieranski}},\ }\href@noop {} {\bibfield  {journal} {\bibinfo  {journal}
  {Physical Review Letters}\ }\textbf {\bibinfo {volume} {45}},\ \bibinfo
  {pages} {569} (\bibinfo {year} {1980})}\BibitemShut {NoStop}%
\bibitem [{\citenamefont {Lin}\ \emph {et~al.}(2003)\citenamefont {Lin},
  \citenamefont {Skaff}, \citenamefont {Emrick}, \citenamefont {Dinsmore},\
  and\ \citenamefont {Russell}}]{lin2003nanoparticle}%
  \BibitemOpen
  \bibfield  {author} {\bibinfo {author} {\bibfnamefont {Y.}~\bibnamefont
  {Lin}}, \bibinfo {author} {\bibfnamefont {H.}~\bibnamefont {Skaff}}, \bibinfo
  {author} {\bibfnamefont {T.}~\bibnamefont {Emrick}}, \bibinfo {author}
  {\bibfnamefont {A.}~\bibnamefont {Dinsmore}}, \ and\ \bibinfo {author}
  {\bibfnamefont {T.~P.}\ \bibnamefont {Russell}},\ }\href@noop {} {\bibfield
  {journal} {\bibinfo  {journal} {Science}\ }\textbf {\bibinfo {volume}
  {299}},\ \bibinfo {pages} {226} (\bibinfo {year} {2003})}\BibitemShut
  {NoStop}%
\bibitem [{\citenamefont {Bresme}\ and\ \citenamefont
  {Oettel}(2007)}]{bresme2007nanoparticles}%
  \BibitemOpen
  \bibfield  {author} {\bibinfo {author} {\bibfnamefont {F.}~\bibnamefont
  {Bresme}}\ and\ \bibinfo {author} {\bibfnamefont {M.}~\bibnamefont
  {Oettel}},\ }\href@noop {} {\bibfield  {journal} {\bibinfo  {journal}
  {Journal of Physics: Condensed Matter}\ }\textbf {\bibinfo {volume} {19}},\
  \bibinfo {pages} {413101} (\bibinfo {year} {2007})}\BibitemShut {NoStop}%
\bibitem [{\citenamefont {Nikolaides}\ \emph {et~al.}(2002)\citenamefont
  {Nikolaides}, \citenamefont {Bausch}, \citenamefont {Hsu}, \citenamefont
  {Dinsmore}, \citenamefont {Brenner}, \citenamefont {Gay},\ and\ \citenamefont
  {Weitz}}]{nikolaides2002electric}%
  \BibitemOpen
  \bibfield  {author} {\bibinfo {author} {\bibfnamefont {M.}~\bibnamefont
  {Nikolaides}}, \bibinfo {author} {\bibfnamefont {A.}~\bibnamefont {Bausch}},
  \bibinfo {author} {\bibfnamefont {M.}~\bibnamefont {Hsu}}, \bibinfo {author}
  {\bibfnamefont {A.}~\bibnamefont {Dinsmore}}, \bibinfo {author}
  {\bibfnamefont {M.}~\bibnamefont {Brenner}}, \bibinfo {author} {\bibfnamefont
  {C.}~\bibnamefont {Gay}}, \ and\ \bibinfo {author} {\bibfnamefont
  {D.}~\bibnamefont {Weitz}},\ }\href@noop {} {\bibfield  {journal} {\bibinfo
  {journal} {Nature}\ }\textbf {\bibinfo {volume} {420}},\ \bibinfo {pages}
  {299} (\bibinfo {year} {2002})}\BibitemShut {NoStop}%
\bibitem [{\citenamefont {Danov}\ \emph {et~al.}(2004)\citenamefont {Danov},
  \citenamefont {Kralchevsky},\ and\ \citenamefont
  {Boneva}}]{danov2004electrodipping}%
  \BibitemOpen
  \bibfield  {author} {\bibinfo {author} {\bibfnamefont {K.~D.}\ \bibnamefont
  {Danov}}, \bibinfo {author} {\bibfnamefont {P.~A.}\ \bibnamefont
  {Kralchevsky}}, \ and\ \bibinfo {author} {\bibfnamefont {M.~P.}\ \bibnamefont
  {Boneva}},\ }\href@noop {} {\bibfield  {journal} {\bibinfo  {journal}
  {Langmuir}\ }\textbf {\bibinfo {volume} {20}},\ \bibinfo {pages} {6139}
  (\bibinfo {year} {2004})}\BibitemShut {NoStop}%
\bibitem [{\citenamefont {Shah}\ \emph {et~al.}(2008)\citenamefont {Shah},
  \citenamefont {Shum}, \citenamefont {Rowat}, \citenamefont {Lee},
  \citenamefont {Agresti}, \citenamefont {Utada}, \citenamefont {Chu},
  \citenamefont {Kim}, \citenamefont {Fernandez-Nieves}, \citenamefont
  {Martinez},\ and\ \citenamefont {Weitz}}]{shah:2008}%
  \BibitemOpen
  \bibfield  {author} {\bibinfo {author} {\bibfnamefont {R.~K.}\ \bibnamefont
  {Shah}}, \bibinfo {author} {\bibfnamefont {H.~C.}\ \bibnamefont {Shum}},
  \bibinfo {author} {\bibfnamefont {A.~C.}\ \bibnamefont {Rowat}}, \bibinfo
  {author} {\bibfnamefont {D.}~\bibnamefont {Lee}}, \bibinfo {author}
  {\bibfnamefont {J.~J.}\ \bibnamefont {Agresti}}, \bibinfo {author}
  {\bibfnamefont {A.~S.}\ \bibnamefont {Utada}}, \bibinfo {author}
  {\bibfnamefont {L.-Y.}\ \bibnamefont {Chu}}, \bibinfo {author} {\bibfnamefont
  {J.-W.}\ \bibnamefont {Kim}}, \bibinfo {author} {\bibfnamefont
  {A.}~\bibnamefont {Fernandez-Nieves}}, \bibinfo {author} {\bibfnamefont
  {C.~J.}\ \bibnamefont {Martinez}}, \ and\ \bibinfo {author} {\bibfnamefont
  {D.~A.}\ \bibnamefont {Weitz}},\ }\href {\doibase
  http://dx.doi.org/10.1016/S1369-7021(08)70053-1} {\bibfield  {journal}
  {\bibinfo  {journal} {Materials Today}\ }\textbf {\bibinfo {volume} {11}},\
  \bibinfo {pages} {18 } (\bibinfo {year} {2008})}\BibitemShut {NoStop}%
\bibitem [{\citenamefont {Dommersnes}\ \emph {et~al.}(2013)\citenamefont
  {Dommersnes}, \citenamefont {Rozynek}, \citenamefont {Mikkelsen},
  \citenamefont {Castberg}, \citenamefont {Kjerstad}, \citenamefont {Hersvik},\
  and\ \citenamefont {Fossum}}]{dommersnes:2013}%
  \BibitemOpen
  \bibfield  {author} {\bibinfo {author} {\bibfnamefont {P.}~\bibnamefont
  {Dommersnes}}, \bibinfo {author} {\bibfnamefont {Z.}~\bibnamefont {Rozynek}},
  \bibinfo {author} {\bibfnamefont {A.}~\bibnamefont {Mikkelsen}}, \bibinfo
  {author} {\bibfnamefont {R.}~\bibnamefont {Castberg}}, \bibinfo {author}
  {\bibfnamefont {K.}~\bibnamefont {Kjerstad}}, \bibinfo {author}
  {\bibfnamefont {K.}~\bibnamefont {Hersvik}}, \ and\ \bibinfo {author}
  {\bibfnamefont {J.~O.}\ \bibnamefont {Fossum}},\ }\href@noop {} {\bibfield
  {journal} {\bibinfo  {journal} {Nature communications}\ }\textbf {\bibinfo
  {volume} {4}} (\bibinfo {year} {2013})}\BibitemShut {NoStop}%
\bibitem [{\citenamefont {Kilic}\ \emph {et~al.}(2007)\citenamefont {Kilic},
  \citenamefont {Bazant},\ and\ \citenamefont {Ajdari}}]{kilic:2007}%
  \BibitemOpen
  \bibfield  {author} {\bibinfo {author} {\bibfnamefont {M.~S.}\ \bibnamefont
  {Kilic}}, \bibinfo {author} {\bibfnamefont {M.~Z.}\ \bibnamefont {Bazant}}, \
  and\ \bibinfo {author} {\bibfnamefont {A.}~\bibnamefont {Ajdari}},\ }\href
  {\doibase 10.1103/PhysRevE.75.021503} {\bibfield  {journal} {\bibinfo
  {journal} {Phys. Rev. E}\ }\textbf {\bibinfo {volume} {75}},\ \bibinfo
  {pages} {021503} (\bibinfo {year} {2007})}\BibitemShut {NoStop}%
\end{thebibliography}%

\end{document}